
\def\ignore#1{}
 

\newcount\sectnum
\newcount\subsectnum
\newcount\eqnumber

\global\eqnumber=1\sectnum=0


\def\lab{(\the\sectnum.\the\eqnumber)}



\def\show#1{#1}



\def\smskip{\vskip 5 pt}
\def\medskip{\vskip 10 pt}
\def\bigskip{\vskip 15 pt}
\def\pn{\par\noindent}
\def\br{\break}

\def\bl{\bigl} 
\def\br{\bigr} 
\def\lf{\left}
\def\ri{\right}

\def\ol#1{\overline{#1}}

\def\a{\alpha}

\def\b{\beta}

\def\m{\mu}
\def\n{\nu}

\def\tl{\tilde}

\def\jstar{J^{\raise0.04pt\hbox{\sevenpoint *}} }
\def\qstar{Q^{\raise0.04pt\hbox{\sevenpoint *}} }

\def\tl{\tilde}

\def\old#1{}
\def\leaderfill{\leaders\hbox to 1em{\hss.\hss}\hfill}


\parindent=2pc
\baselineskip=15pt
\vsize=8.7 true in
\voffset=0.125 true in
\parskip=3pt


\def\minprob#1#2#3{$$\eqalign{&\hbox{minimize\ \ }#1\cr &\hbox{subject to\ \
}#2\cr}\ifnum 0=#3{}\else\eqno(#3)\fi$$}        
     
\def\maxprob#1#2#3{$$\eqalign{&\hbox{maximize\ \ }#1\cr &\hbox{subject to\ \
}#2\cr}\ifnum 0=#3{}\else\eqno(#3)\fi$$}        
     
\def\aligntwo#1#2#3#4#5{$$\eqalign{#1&#2\cr #3&#4\cr}
\ifnum 0=#5{}\else\eqno(#5)\fi$$}
\def\alignthree#1#2#3#4#5#6#7{$$\eqalign{#1&#2\cr #3&#4\cr #5&#6\cr}
\ifnum 0=#7{}\else\eqno(#7)\fi$$}


\def\eqnum{\eqno{\hbox{(\the\sectnum.\the\eqnumber)}\global\advance\eqnumber
by1}}

\def\eqnu{\eqno{\hbox{(\the\sectnum.\the\eqnumber)}\global\advance\eqnumber
by1}}

\newcount\examplnumber
\def\examplnum{\global\advance\examplnumber by1}

\newcount\figrnumber
\def\figrnum{\global\advance\figrnumber by1}

\newcount\propnumber
\def\propnum{\global\advance\propnumber by1}

\newcount\defnumber
\def\defnum{\global\advance\defnumber by1}

\newcount\lemmanumber
\def\lemmanum{\global\advance\lemmanumber by1}

\newcount\assumptionnumber
\def\assumptionnum{\global\advance\assumptionnumber by1}

\def\exampl{\the\sectnum.\the\examplnumber}
\def\figr{\the\sectnum.\the\figrnumber}
\def\propn{\the\sectnum.\the\propnumber}
\def\defn{\the\sectnum.\the\defnumber}
\def\lemman{\the\sectnum.\the\lemmanumber}
\def\assumptionn{\the\sectnum.\the\assumptionnumber}

\def\section#1{\goodbreak\vskip 3pc plus 6pt minus 3pt\leftskip=-2pc
   \global\advance\sectnum by 1\eqnumber=1
\global\examplnumber=1\figrnumber=1\propnumber=1\defnumber=1\lemmanumber=1\assumptionnumber=1%
   \line{\hfuzz=1pc{\hbox to 3pc{\bf 
   \vtop{\hfuzz=1pc\hsize=38pc\hyphenpenalty=10000\noindent\uppercase{\the\sectnum.\quad #1}}\hss}}
			\hfill}
			\leftskip=0pc\nobreak\tenf
			\vskip 1pc plus 4pt minus 2pt\noindent\ignorespaces}



\def\sect#1{\noindent\leftskip=-2pc\tenf
   \goodbreak\vskip 1pc plus 4pt minus 2pt
                \global\advance\subsectnum by 1\eqnumber=1
   \line{\hfuzz=1pc{\hbox to 3pc{\bf 
   \vtop{\hfuzz=1pc\hsize=38pc\hyphenpenalty=10000\noindent\uppercase{{\bf #1}}}\hss}}
                        \hfill}
   \leftskip=0pc\nobreak\tenf
                        \vskip 1pc plus 4pt minus 2pt\nobreak\noindent\ignorespaces}

\def\subsection#1{\noindent\leftskip=0pc\tenf
   \goodbreak\vskip 1pc plus 4pt minus 2pt
   \line{\hfuzz=1pc{\hbox to 3pc{\bf 
   \vtop{\hfuzz=1pc\hsize=38pc\hyphenpenalty=10000\noindent{\bf #1}}\hss}}
                        \hfill}
   \leftskip=0pc\nobreak\tenf
                        \vskip 1pc plus 4pt minus 2pt\nobreak\noindent\ignorespaces}
\def\subsubsection#1{\goodbreak\vskip 1pc plus 4pt minus 2pt
   \hfuzz=3pc\leftskip=0pc\noindent\tenit #1 \nobreak\tenf\vskip 6pt plus 1pt
                                minus 1pt\nobreak\ignorespaces\leftskip=0pc}
%

\def\beginexample#1{\noindent\goodbreak\vskip 6pt plus 1pt minus 1pt
\noindent
  \hbox {\bf Example #1\hss}
  \nobreak\vskip 4pt plus 1pt minus 1pt \nobreak\noindent\ninef
  \global\advance
                \leftskip by\parindent\pn}
\def\endexample{\vskip 12pt\tenf\par
  \global\advance\leftskip by -\parindent
  }

\def\beginexercise#1{\noindent\goodbreak\vskip 6pt plus 1pt minus 1pt \noindent\global\normalbaselineskip=12pt
  \hbox {\bf Exercise #1\hss}
  \nobreak\vskip 4pt plus 1pt minus 1pt 
  \nobreak\noindent\ninef\global\advance\leftskip
                        by\parindent\pn}
\def\endexercise{\vskip 12pt\tenf\par
  \global\advance\leftskip by -\parindent
  }

\def\beginsection#1{\noindent\goodbreak\vskip 6pt plus 1pt minus 1pt \noindent\global\normalbaselineskip=12pt
  \hbox {\it #1\hss}
  \vskip 0.1pt plus 1pt minus 1pt \nobreak\noindent\ninef\global\advance
                \leftskip by\parindent\noindent\pn}
\def\endsection{\vskip 12pt\tenf\par
  \global\advance\leftskip by -\parindent
}

%


\def\proposition#1{\smskip\pn{\bf Proposition #1}\quad}
\def\proof{\smskip\pn{\bf Proof:}\quad}

 \def\qed{\quad{\bf
Q.E.D.} \par\bigskip}
\def\ref{\smskip\pn}

\def\chapter#1#2{{\bf \centerline{\helbigbig
{#1}}}\bigskip\bigskip{\bf \centerline{\helbigbig
{#2}}}\bigskip\bigskip} 



\def\longpapertitle#1#2#3{{\bf \centerline{\helbigb
{#1}}}\bigskip{\bf \centerline{\helbigb
{#2}}}\bigskip\bigskip{\centerline{
by}}\bigskip{\bf \centerline{
{#3}}}\bigskip\bigskip} 


\def\nitem#1{\smskip\item{#1}}

\newcount\alphanum
\newcount\romnum

\def\alphaenumerate{\ifcase\alphanum \or (a)\or (b)\or (c)\or (d)\or (e)\or
(f)\or (g)\or (h)\or (i)\or (j)\or (k)\fi}
\def\romenumerate{\ifcase\romnum \or (i)\or (ii)\or (iii)\or (iv)\or (v)\or
(vi)\or (vii)\or (viii)\or (ix)\or (x)\or (xi)\fi}

\def\alist{\begingroup\vskip10pt\alphanum=1
\parskip=2pt\parindent=0pt \leftskip=3pc
\everypar{\llap{{\rm\alphaenumerate\hskip1em}}\advance\alphanum by1}}

\def\nolist{\begingroup\vskip10pt\alphanum=0
\parskip=2pt\parindent=0pt \leftskip=3pc
\everypar{\llap{\global\advance\alphanum by1(\the\alphanum)\hskip1em}}}

\def\romlist{\begingroup\vskip10pt\romnum=1
\parskip=2pt\parindent=0pt \leftskip=5pc
\everypar{\llap{{\rm\romenumerate\hskip1em}}\advance\romnum by1}}



\long\def\fig#1#2#3{\vbox{\vskip1pc\vskip#1
\prevdepth=12pt \baselineskip=12pt
\vskip1pc
\hbox to\hsize{\hfill\vtop{\hsize=25pc\noindent{\eightbf Figure #2\ }
{\eightpoint#3}}\hfill}}}

\long\def\widefig#1#2#3{\vbox{\vskip1pc\vskip#1
\prevdepth=12pt \baselineskip=12pt
\vskip1pc
\hbox to\hsize{\hfill\vtop{\hsize=28pc\noindent{\eightbf Figure #2\ }
{\eightpoint#3}}\hfill}}}

\long\def\table#1#2{\vbox{\vskip0.5pc
\prevdepth=12pt \baselineskip=12pt
\hbox to\hsize{\hfill\vtop{\hsize=25pc\noindent{\eightbf Table #1\ }
{\eightpoint#2}}\hfill}}}

 
\def\rightheadline#1{\headline{\tenrm\hfil #1}}


\long\def\leftfig#1#2{\vbox{\smskip\hsize=220pt
\vtop{{\noindent {\bf #1}}}
\smskip
\noindent
\vbox{{\noindent #2}}
}}

\long\def\rightfig#1#2#3{\vbox{\smskip\vskip#1
\prevdepth=12pt \baselineskip=12pt
\hsize=210pt
\smskip
\vbox{\noindent{\eightbold #2}
\hskip1em{\eightpoint#3}}
}}

\long\def\concept#1#2#3#4#5{\bigskip\hrule
\vbox{\hbox{\leftfig{#1}{#2} \hskip3em
\rightfig{#3}{#4}{#5}} \smskip}
\hrule\bigskip}


\long\def\bconcept#1#2#3#4#5#6#7{
\vbox{
\hbox to \hsize{\vtop{\par #1}}
\concept{#2}{#3}{#4}{#5}{#6}
\hbox to \hsize{\vtop{\par #7}}
\smskip}
}




\def\boxit#1{\vbox{\hrule\hbox{\vrule\kern3pt
                                \vbox{\kern3pt#1\kern3pt}\kern3pt\vrule}\hrule}}
\def\centerboxit#1{$$\vbox{\hrule\hbox{\vrule\kern3pt
                                \vbox{\kern3pt#1\kern3pt}\kern3pt\vrule}\hrule}$$}

\long\def\boxtext#1#2{$$\boxit{\vbox{\hsize #1\noindent\strut #2\strut}}$$}

%
%
%

\def\picture #1 by #2 (#3){
  \vbox to #2{
    \hrule width #1 height 0pt depth 0pt
    \vfill
    \special{picture #3} 
    }
  }

\def\scaledpicture #1 by #2 (#3 scaled #4){{
  \dimen0=#1 \dimen1=#2
  \divide\dimen0 by 1000 \multiply\dimen0 by #4
  \divide\dimen1 by 1000 \multiply\dimen1 by #4
  \picture \dimen0 by \dimen1 (#3 scaled #4)}
  }

%
%

\long\def\captfig#1#2#3#4#5{\vbox{\vskip1pc
\hbox to\hsize{\hfill{\picture #1 by #2 (#3)}\hfill}
\prevdepth=9pt \baselineskip=9pt
\vskip1pc
\hbox to\hsize{\hfill\vtop{\hsize=24pc\noindent{\eightbold Figure #4}
\hskip1em{\eightpoint#5}}\hfill}}}

%
%
%

\def\illustration #1 by #2 (#3){
  \vskip#2\hskip#1\special{illustration #3} 
    }

\def\scaledillustration #1 by #2 (#3 scaled #4){{
  \dimen0=#1 \dimen1=#2
  \divide\dimen0 by 1000 \multiply\dimen0 by #4
  \divide\dimen1 by 1000 \multiply\dimen1 by #4
  \illustration \dimen0 by \dimen1 (#3 scaled #4)}
  }


\newbox\graybox
\newdimen\xgrayspace
\newdimen\ygrayspace
%
%
%
%
%
%
%
%
%

\def\Textshade#1#2#3#4#5#6{%
    \xgrayspace=#4pt%
    \ygrayspace=#4pt%
    \def\grayshade{#3}%
    \def\linewidth{#5}%
    \def\theradius{#6}%
    \setbox\graybox=\hbox{\surroundboxa{#2}}%
    \hbox{%
    \hbox to 0pt{%
    \PScommands
}%
    \box\graybox}}%
%
%
\long%

\long%
\def\Parashade#1#2#3#4#5#6#7{%
    \xgrayspace=#4pt%
    \ygrayspace=#4pt%
    \def\grayshade{#3}%
    \def\linewidth{#5}%
    \def\theradius{#6}%
    \def\thevskip{#7pt}%
    \setbox\graybox=\hbox{\surroundboxb{#2}}%
    \vskip\thevskip%
    \hbox{%
    \hbox to 0pt{%
    \PScommands
}%
     \box\graybox}%
     \vskip\thevskip%
}%
%
%
%
\long\def\surroundboxa#1{\leavevmode\hbox{\vtop{%
\vbox{\kern\ygrayspace%
\hbox{\kern\xgrayspace#1%
      \kern\xgrayspace}}\kern\ygrayspace}}}
%
%
\long\def\surroundboxb#1{\leavevmode\hbox{\vtop{%
\vbox{\kern\ygrayspace%
\hbox{\kern\xgrayspace\vbox{\advance\hsize-2\xgrayspace#1}%
      \kern\xgrayspace}}\kern\ygrayspace}}}
%
%
%
\long\def\PScommands{%
\special{rawpostscript
/sharpbox{%
           newpath
           xmin ymin moveto
           xmin ymax lineto
           xmax ymax lineto
           xmax ymin lineto
           xmin ymin lineto
           closepath 
          }bind def
}%
\special{rawpostscript
/sharpboxnb{%
           newpath
           xmin ymin moveto
           xmin ymax lineto
           xmax ymax lineto
           xmax ymin lineto
          }bind def
}%
\special{rawpostscript
/sharpboxnt{%
           newpath
           xmin ymax moveto
           xmin ymin lineto
           xmax ymin lineto
           xmax ymax lineto
          }bind def
}%
\special{rawpostscript
/roundbox{%
           newpath
           xmin radius add ymin moveto
           xmax ymin xmax ymax radius arcto
           xmax ymax xmin ymax radius arcto
           xmin ymax xmin ymin radius arcto
           xmin ymin xmax ymin radius arcto 16 {pop} repeat
           closepath
          }bind def
}%
\special{rawpostscript
/sharpcorners{%
               sharpbox gsave grayshade setgray fill grestore 
               linewidth setlinewidth stroke
              }bind def
}%
\special{rawpostscript
/sharpcornersnt{%
               sharpboxnt gsave grayshade setgray fill grestore 
               linewidth setlinewidth stroke
              }bind def
}%
\special{rawpostscript
/sharpcornersnb{%
               sharpboxnb gsave grayshade setgray fill grestore 
               linewidth setlinewidth stroke
              }bind def
}%
\special{rawpostscript
/roundcorners{%
               roundbox gsave grayshade setgray fill grestore 
               linewidth setlinewidth stroke
              }bind def
}%
\special{rawpostscript
/plainbox{%
           sharpbox grayshade setgray fill 
          }bind def
}%
%
\special{rawpostscript
/roundnoframe{%
               roundbox grayshade setgray fill 
              }bind def
}%
\special{rawpostscript
/sharpnoframe{%
               sharpbox grayshade setgray fill 
              }bind def
}%
}%
%
%

\def\pshade#1{\Parashade{sharpcorners}{#1}{0.95}{10}{0.5}{10}{10}}


\def\boxit#1{\vbox{\hrule\hbox{\vrule\kern3pt
                                \vbox{\kern3pt#1\kern3pt}\kern3pt\vrule}\hrule}}

\def\boxitnb#1{\vbox{\hrule\hbox{\vrule\kern3pt
                                \vbox{\kern3pt#1\kern3pt}\kern3pt\vrule}}}

\def\boxitnt#1{\vbox{\hbox{\vrule\kern3pt
                                \vbox{\kern3pt#1\kern3pt}\kern3pt\vrule}\hrule}}

\long\def\boxtext#1#2{$$\boxit{\vbox{\hsize #1\noindent\strut #2\strut}}$$}
\long\def\boxtextnb#1#2{$$\boxitnb{\vbox{\hsize #1\noindent\strut #2\strut}}$$}
\long\def\boxtextnt#1#2{$$\boxitnt{\vbox{\hsize #1\noindent\strut #2\strut}}$$}

\def\texshopbox#1{\boxtext{462pt}{\vskip-1.5pc\pshade{\vskip-1.0pc#1\vskip-2.0pc}}}
\def\texshopboxnt#1{\boxtextnt{462pt}{\vskip-1.5pc\pshade{\vskip-1.0pc#1\vskip-2.0pc}}}
\def\texshopboxnb#1{\boxtextnb{462pt}{\vskip-1.5pc\pshade{\vskip-1.0pc#1\vskip-2.0pc}}}


%
%
%
%
%
%
%
%
\font\helbigbig=cmr10 scaled 2500%
\font\helbigb=cmbx10 scaled 1500%
\font\eightbold=cmbx8%

\def\tenf{\hel}%
\def\tenit{\heli}%
\def\ninef{\ninehel}%
\def\nineit{\nineheli}%
%
%


\font\tenrm=cmr10%
\font\teni=cmmi10%
\font\tensy=cmsy10%
\font\tenbf=cmbx10%
\font\tentt=cmtt10%
\font\tenit=cmti10%
\font\tensl=cmsl10%

\def\tenpoint{\def\rm{\fam0\tenrm}%
\textfont0=\tenrm%
\textfont1=\teni%
\textfont2=\tensy%
\textfont\itfam=\tenit%
\textfont\slfam=\tensl%
\textfont\ttfam=\tentt%
\textfont\bffam=\tenbf%
\scriptfont0=\sevenrm%
\scriptfont1=\seveni%
\scriptfont2=\sevensy%
\scriptscriptfont0=\sixrm%
\scriptscriptfont1=\sixi%
\scriptscriptfont2=\sixsy%
\def\it{\fam\itfam\tenit}%
\def\tt{\fam\ttfam\tentt}%
\def\sl{\fam\slfam\tensl}%
\scriptfont\bffam=\sevenbf%
\scriptscriptfont\bffam=\sixbf%
\def\bf{\fam\bffam\tenbf}%
\normalbaselineskip=18pt%
\normalbaselines\rm}%

\font\ninerm=cmr9%
\font\ninebf=cmbx9%
\font\nineit=cmti9%
\font\ninesy=cmsy9%
\font\ninei=cmmi9%
\font\ninett=cmtt9%
\font\ninesl=cmsl9%

\def\ninepoint{\def\rm{\fam0\ninerm}%
\textfont0=\ninerm%
\textfont1=\ninei%
\textfont2=\ninesy%
\textfont\itfam=\nineit%
\textfont\slfam=\ninesl%
\textfont\ttfam=\ninett%
\textfont\bffam=\ninebf%
\scriptfont0=\sixrm%
\scriptfont1=\sixi%
\scriptfont2=\sixsy%
\def\it{\fam\itfam\nineit}%
\def\tt{\fam\ttfam\ninett}%
\def\sl{\fam\slfam\ninesl}%
\scriptfont\bffam=\sixbf%
\scriptscriptfont\bffam=\fivebf%
\def\bf{\fam\bffam\ninebf}%
\normalbaselineskip=16pt%
\normalbaselines\rm}%

\font\eightrm=cmr8%
\font\eighti=cmmi8%
\font\eightsy=cmsy8%
\font\eightbf=cmbx8%
\font\eighttt=cmtt8%
\font\eightit=cmti8%
\font\eightsl=cmsl8%

\def\eightpoint{\def\rm{\fam0\eightrm}%
\textfont0=\eightrm%
\textfont1=\eighti%
\textfont2=\eightsy%
\textfont\itfam=\eightit%
\textfont\slfam=\eightsl%
\textfont\ttfam=\eighttt%
\textfont\bffam=\eightbf%
\scriptfont0=\sixrm%
\scriptfont1=\sixi%
\scriptfont2=\sixsy%
\scriptscriptfont0=\fiverm%
\scriptscriptfont1=\fivei%
\scriptscriptfont2=\fivesy%
\def\it{\fam\itfam\eightit}%
\def\tt{\fam\ttfam\eighttt}%
\def\sl{\fam\slfam\eightsl}%
\scriptscriptfont\bffam=\fivebf%
\def\bf{\fam\bffam\eightbf}%
\normalbaselineskip=14pt%
\normalbaselines\rm}%

\font\sevenrm=cmr7%
\font\seveni=cmmi7%
\font\sevensy=cmsy7%
\font\sevenbf=cmbx7%

\def\sevenpoint{%
   \def\rm{\sevenrm}\def\bf{\sevenbf}%
   \def\smc{\sevensmc}\baselineskip=12pt\rm}%

\font\sixrm=cmr6%
\font\sixi=cmmi6%
\font\sixsy=cmsy6%
\font\sixbf=cmbx6%

\fontdimen13\tensy=2.6pt%
\fontdimen14\tensy=2.6pt%
\fontdimen15\tensy=2.6pt%
\fontdimen16\tensy=1.2pt%
\fontdimen17\tensy=1.2pt%
\fontdimen18\tensy=1.2pt%

\def\tenf{\tenpoint}%
\def\ninef{\ninepoint}%
%



\def\texshopbox#1{\boxtext{462pt}{\vskip-1.5pc\pshade{\vskip-1.0pc#1\vskip-2.0pc}}}
\def\texshopboxnt#1{\boxtextnt{462pt}{\vskip-1.5pc\pshade{\vskip-1.0pc#1\vskip-2.0pc}}}
\def\texshopboxnb#1{\boxtextnb{462pt}{\vskip-1.5pc\pshade{\vskip-1.0pc#1\vskip-2.0pc}}}

\long\def\fig#1#2#3{\vbox{\vskip1pc\vskip#1
\prevdepth=12pt \baselineskip=12pt
\vskip1pc
\hbox to\hsize{\hfill\vtop{\hsize=30pc\noindent{\eightbf Figure #2\ }
{\eightpoint#3}}\hfill}}}

\def\show#1{}

\def\frac#1#2{{#1\over #2}}

\rightheadline{\botmark}

\pageno=1





%

%


\input epsf




\pn {\bf July 2021 (revised October 2021)}\hfill{\bf ASU/SCAI Report}
\bigskip \bigskip

\bigskip\bigskip\bigskip

\def\longpapertitle#1#2#3{{\bf \centerline{\helbigb
{#1}}}\medskip{\bf \centerline{\helbigb
{#2}}}\medskip{\centerline{
by}}\medskip{\bf \centerline{
{#3}}}\bigskip}

\longpapertitle{Distributed Asynchronous Policy Iteration for}{Sequential Zero-Sum Games and Minimax Control}{{Dimitri Bertsekas\footnote{\dag}{\ninepoint Fulton Professor of Computational Decision Making, School of Computing, Informatics, and Decision Systems Engineering, Arizona State University, Tempe, AZ.}}}

\centerline{\bf Abstract}

We introduce a contractive abstract dynamic programming framework and related policy iteration algorithms, specifically designed for sequential zero-sum games and minimax problems with a general structure. Aside from greater generality, the  advantage of our algorithms over alternatives is that they resolve some long-standing convergence difficulties of the ``natural" policy iteration algorithm, which have been known since the Pollatschek and Avi-Itzhak method [PoA69] for finite-state Markov games.  Mathematically, this ``natural" algorithm is a form of Newton's method for solving Bellman's equation, but Newton's method, contrary to the case of single-player DP problems, is not globally convergent in the case of a minimax problem, because the Bellman operator may have components that are neither convex nor concave. Our algorithms address this difficulty by introducing alternating player choices, and by using a policy-dependent mapping with a uniform sup-norm contraction property, similar to earlier works by Bertsekas and Yu [BeY10], [BeY12], [YuB13]. Moreover, our algorithms allow a convergent and highly parallelizable implementation, which is based on state space partitioning, and distributed asynchronous policy evaluation and policy improvement operations within each set of the partition. Our framework is also suitable for the use of reinforcement learning methods based on aggregation, which may be useful for large-scale problem instances.

\vskip-2pc
\vfill\eject
\section{Introduction}
\mark{Introduction}

\vskip-2pc

\pn The purpose of this paper is to discuss abstract dynamic programming (DP) frameworks and policy iteration (PI) methods for sequential minimax problems. In addition to being more efficient and reliable than alternatives, our methods are well suited for distributed asynchronous implementation. In Sections 1 and 2, we will discuss an abstract DP framework, which is well-known. We will revisit abstract PI algorithms within this framework and show how they relate to known algorithms for minimax control. We will also discuss how these algorithms when applied to discounted and terminating zero-sum Markov games, lead to methods such as the ones by Hoffman and Karp [HoK66], and by Pollatschek and Avi-Itzhak [PoA69]. Related methods have been discussed for Markov games by van der Wal  [Van78], Tolwinski [Tol89], Filar and Tolwinski [FiT91],  Filar and Vrieze [FiV97], and for stochastic shortest gamess, by Patek and Bertsekas [PaB99], and Yu [Yu14]; see also Perolat et al.\ [PSP15], [PPG16], and the survey by Zhang, Yang, and Basar [ZYB21] for related reinforcement learning methods. We will note some of the drawbacks of these algorithms, particularly the need to solve a substantial optimization problem as part of the policy evaluation phase. These drawbacks motivate new PI algorithms and a different abstract framework, based on an alternating player choices format, which we will introduce in Section 3.

In our initial problem formulation, the focus of Sections 1 and 2, 
we consider abstract sequential infinite horizon zero-sum game and minimax problems, which involve two players that choose controls at each  state $x$ of some state space $X$, from within some state-dependent constraint sets: a {\it minimizer\/}, who selects a control $u$ from within a subset $U(x)$ of a control space $U$, and a {\it maximizer\/},  who selects a control $v$ from within a subset $V(x)$ of a control space $V$.  The spaces $X$, $U$, and $V$ are arbitrary.
Functions $\m:X\mapsto U$ and $\n:X\mapsto V$ such that $\m(x)\in U(x)$ and $\n(x)\in V(x)$ for all $x\in X$, are called {\it policies} for the minimizer and the maximizer, respectively. The set of policies for the minimizer and the maximizer are denoted by ${\cal M}$ and ${\cal N}$, respectively.

The main idea of abstract DP formulations is to start with a general mapping that defines the Bellman equation of the problem, also referred to generically as {\it Bellman operator} in this paper. Special cases of this operator define a variety of deterministic optimal control problems, Markovian decision problems with additive and risk-sensitive cost functions, minimax and zero-sum game problems, and others. In this paper we focus on minimax problems. In particular, we introduce a suitable real-valued mapping 
$$H(x,u,v,J),\qquad x\in X,\ u\in U(x),\ v\in V(x),\ J\in B(X),\xdef\firstmap{\lab}\eqnum\show{oneo}$$
(a different mapping, which separates the choices of the two players, will be given in Section 3). In Eq.\ \firstmap, $B(X)$ is the space of real-valued functions on $X$ that are bounded with respect to a weighted sup-norm
$$\|J\|=\sup_{x\in X}{\big|J(x)\big|\over \xi(x)},\qquad J\in B(X),\xdef\supnorm{\lab}\eqnum\show{oneo}$$
where $\xi$ is a function taking a positive value $\xi(x)$ for each $x\in X$.
Our main assumption is the following:

\texshopbox{\pn{\bf Assumption 1.1: (Contraction Assumption)} For every  $\m\in{\cal M}$, $\n\in{\cal N}$, consider the operator $T_{\m,\n}$ that maps a function $J\in B(X)$ to the function $T_{\m,\n}J$ defined by
$$(T_{\m,\n}J)(x)=H\big(x,\m(x),\n(x),J\big),\qquad x\in X.\eqnum\show{oneo}$$
We assume the following:
\nitem{(a)} $T_{\m,\n}J$ belongs to $B(X)$ for all $J\in B(X)$.
\nitem{(b)} There exists an $\a\in(0,1)$ such that  for all  $\m\in{\cal M}$, $\n\in{\cal N}$, the operator $T_{\m,\n}$ is a contraction mapping  of modulus $\a$ with respect to the weighted sup-norm \supnorm, i.e., for all $J,J'\in B(X)$, $\m\in{\cal M}$, and  $\n\in{\cal N}$,
$$\|T_{\m,\n}J-T_{\m,\n}J'\|=\sup_{x\in X}{\big|(T_{\m,\n}J)(x)-(T_{\m,\n}J')(x)\big|\over \xi(x)}\le \a\|J-J'\|.$$
}

Since $T_{\m,\n}$ is a contraction within the complete space $B(X)$, under the preceding assumption, it has a unique fixed point $J_{\m,\n}\in B(X)$. 
We are interested in the operator $T:B(X)\mapsto B(X)$, defined by
$$(TJ)(x)=\inf_{u\in U(x)}\sup_{v\in V(x)}H(x,u,v,J),\qquad x\in X,\xdef\minimaxt{\lab}\eqnum\show{oneo}$$
or equivalently,
$$(TJ)(x)=\inf_{\m\in{\cal M}}\sup_{\n\in{\cal N}}(T_{\m,\n}J)(x),\qquad x\in X.\xdef\minimaxtalt{\lab}\eqnum\show{oneo}$$
An important fact is that {\it $T$ is a contraction mapping from $B(X)$ to $B(X)$\/}.  Indeed from Assumption 1.1(b), we have for all $x\in X$, $\m\in{\cal M}$, and $\n\in{\cal N}$,
$$(T_{\m,\n}J)(x)\le (T_{\m,\n}J')(x)+\a\|J-J'\|\,\xi(x).$$
Taking the supremum over  $\n\in{\cal N}$ of both sides above, and then the infimum over $\m\in{\cal M}$, and using Eq.\ \minimaxtalt, we obtain
$$(TJ)(x)\le (TJ')(x)+\a\|J-J'\|\,\xi(x),\qquad \hbox{for all }x\in X.$$
Similarly, by reversing the roles of $J$ and $J'$, we obtain
$$(TJ')(x)\le (TJ')(x)+\a\|J-J'\|\,\xi(x),\qquad \hbox{for all }x\in X.$$
Combining the preceding two relations, we have
$$\big|(TJ)(x)-(TJ')(x)\big|\le \a\|J-J'\|\,\xi(x),\qquad \hbox{for all }x\in X,$$
and by dividing with $\xi(x)$, and taking supremum over $x\in X$, it follows that
$$\|TJ-TJ'\|\le \a\|J-J'\|.$$
Thus $T$ is a contraction mapping from $B(X)$ to $B(X)$, with respect to the sup-norm \supnorm, with modulus $\a$, and has a unique fixed point within $B(X)$, which we denote by $\jstar$.
 
\subsubsection{Bellman's Equation and Minimax Optimal Policies}

\pn Given a mapping $H$ of the form \firstmap\ that satisfies Assumption 1.1, we are interested in computing  the fixed point $\jstar$ of $T$, i.e., a function $\jstar$ such that
$$\jstar(x)=\inf_{u\in U(x)}\sup_{v\in V(x)}H(x,u,v,\jstar),\qquad \hbox{for all }x\in X.\xdef\belfixedpoint{\lab}\eqnum\show{oneo}$$
Moreover, we are interested in finding a policy $\m^*\in{\cal M}$ (if it exists) that attains the infimum for all $x\in X$ as in the following equation
$$\m^*(x)\in\arg\min_{u\in U(x)} \ol H(x,u,\jstar),\qquad \hbox{for all }x\in X,$$ 
where for all $x\in X$, $u\in U(x)$, and $J\in B(X)$, the mapping $\ol H$ is defined by
$$\ol H(x,u,J)=\sup_{v\in V(x)}H(x,u,v,J).$$ 
We are also interested in finding a  policy $\n^*\in{\cal N}$ (if it exists) that attains the supremum for all $x\in X$ as in the following equation
$$\n^*(x)\in\arg\max_{v\in V(x)} H\big(x,\m^*(x),v,\jstar\big),\qquad \hbox{for all }x\in X.$$

In the context of a sequential minimax problem, which is addressed by DP,  the fixed point equation $\jstar=T\jstar$ is viewed as a form of Bellman's equation. In this case, $\jstar(x)$ is the minimax cost starting from state $x$. Moreover {\it $\m^*$ is an optimal policy for the minimizer in a minimax sense, while $\n^*$ is a corresponding worst case response of the maximizer\/}. Under suitable assumptions on $H$ (such as convexity in $u$ and concavity in $v$) the order of minimization and maximization can be interchanged in the preceding relations, in which case it can be shown that $(\m^*,\n^*)$ is a saddle point (within the space ${\cal M}\times{\cal N}$) of the minimax value $J_{\m,\n}(x)$, for every $x\in X$. 

\old{
Special cases of the mapping $H$ and the associated fixed point problems have been studied extensively in the single-player case, where there is no maximizer; see the discussion of Section 2.
Generally, in DP applications, the mapping $H$ typically has a monotonicity property, whereby for any two functions $J\in B(X)$ and $J'\in B(X)$ with $J(x)\le J'(x)$, for all $x\in X$, we have
$$H(x,u,v,J)\le H(x,u,v,J'),\qquad\hbox{for all }x\in X,\ u\in U(x),\ v\in V(x).$$
The PI convergence results to be quoted in the next section require this monotonicity structure.
}

\subsubsection{Markov Games}

\pn The simplest special case of a sequential stochastic game problem, which relates to our abstract framework, was introduced in the paper by Shapley [Sha53] for undiscounted finite-state problems, with a termination state, where the Bellman operator $T_{\m,\n}$ is contractive with respect to the sup-norm for all $\m\in{\cal M}$, and $\n\in{\cal N}$. Shapley's work brought the contraction mapping approach to prominence in DP and sequential game analysis, and was subsequently extended by several authors in both undiscounted and discounted settings; see e.g., the book by Filar and Vrieze [FiV97], the lecture notes by Kallenberg [Kal20], and the works referenced there. 
 Let us now describe the finite-state zero-sum game problems that descend from Shapley's work, and are often called ``Markov games" (the name was introduced by Zachrisson [Zac64]).

\xdef\exampleone{\exampl}\examplnum\show{myexample}

\beginexample{\exampleone\ (Discounted Finite-State Markov Games)}\pn Consider two players that play repeated matrix games at each of an infinite number of stages, using mixed strategies. The game played at a given stage is defined by a state $x$ that takes values in a finite set $X$, and changes from one stage to the next according to a Markov chain whose transition probabilities are influenced by the players' choices. At each stage and state $x\in X$, the minimizer selects
a probability distribution
$u=(u_1,\ldots ,u_n)$ over 
$n$ possible choices $i=1,\ldots,n$, and  the maximizer selects a
probability distribution $v=(v_1,\ldots ,v_m)$ over $m$ possible choices $j=1,\ldots,m$. 
If the minimizer chooses $i$ and the maximizer chooses $j$, the payoff of the stage is $a_{ij}(x)$ and depends on the state $x$. Thus the
expected payoff of the stage is
$\sum_{i,j}a_{ij}(x)u_iv_j$ or $u'A(x)v$, where $A(x)$ is the $n\times m$ matrix with
components $a_{ij}(x)$ ($u$ and $v$ are viewed as column vectors, and a prime denotes transposition).

The state evolves according to transition probabilities $q_{xy}(i,j)$, where $i$ and $j$ are the moves selected by the minimizer and the maximizer, respectively (here $y$ represents the next state and game to be played after moves $i$ and $j$ are chosen at the game represented by $x$). When the state is $x$, under $u$ and $v$, the  state transition probabilities are
$$p_{xy}(u,v)=\sum_{i=1}^n\sum_{j=1}^mu_iv_jq_{xy}(i,j)=u'Q_{xy}v,$$
where $Q_{xy}$ is the $n\times m$  matrix that has components $q_{xy}(i,j)$.
Payoffs are discounted by $\a\in(0,1)$, and the objectives of the minimizer and maximizer, are to minimize and to maximize the total discounted expected payoff, respectively.

As shown by Shapley [Sha53], the problem can be formulated as a fixed point problem involving the mapping $H$ given by
$$H(x,u,v,J)=u'A(x)v+\a\sum_{y\in X}p_{xy}(u,v)J(y)=u'\lf(A(x)+\a \sum_{y\in X}Q_{xy}J(y)\ri)v.\xdef\gamemap{\lab}\eqnum\show{oneo}$$
It can be verified that $H$ satisfies the contraction Assumption 1.1 [with $\xi(x)\equiv1$]. Thus the corresponding operator $T$ is an unweighted sup-norm contraction, and its unique fixed point $J^*$ satisfies the Bellman equation
$$J^*(x)=(TJ^*)(x)=\min_{u\in U}\max_{v\in V}H(x,u,v,J^*),\qquad \hbox{for all } x\in X,\xdef\bellmanmarkov{\lab}\eqnum\show{oneo}$$
where $U$ and $V$ denote the sets of probability distributions $u=(u_1,\ldots,u_n)$ and $v=(v_1,\ldots ,v_m)$, respectively. 

Since the matrix defining the mapping $H$ of Eq.\ \gamemap,
$$A(x)+\a \sum_{y\in X}Q_{xy}J(y),$$ 
is independent of $u$ and $v$, we may view $J^*(x)$ as the value of a static (nonsequential) matrix game that depends on $x$.
In particular, from a fundamental saddle point theorem for matrix games, we have
$$\min_{u\in U}\max_{v\in V}H(x,u,v,J^*)=\max_{v\in V}\min_{u\in U}H(x,u,v,J^* ),\qquad \hbox{for all } x\in X.\xdef\saddlemdp{\lab}\eqnum\show{oneo}$$
 It was shown by Shapley [Sha53] that the strategies obtained by solving the static saddle point problem \saddlemdp\ correspond to a saddle point of the sequential game in the space of strategies. Thus once we find $J^*$ as the fixed point of the mapping $T$ [cf.\ Eq.\ \bellmanmarkov], we can obtain equilibrium policies for the minimizer and maximizer by solving the matrix game \saddlemdp.  
  
\endexample

\vskip-1pc

\xdef\exampletwo{\exampl}\examplnum\show{myexample}

\beginexample{\exampletwo\ (Undiscounted Finite-State Markov Games with a Termination State)}\pn Here the problem is the same as in the preceding example, except that there is no discount factor ($\a=1$), and in addition to the states in $X$, there is a termination state $t$ that is cost-free and absorbing. In this case the mapping $H$ is given by
$$H(x,u,v,J)=u'\lf(A(x)+\sum_{y\in X}Q_{xy}J(y)\ri)v,\xdef\tmapminimaxone{\lab}\eqnum\show{oneo}$$
cf.\ Eq.\ \gamemap, where the matrix of transition probabilities $Q_{xy}$ may be substochastic, while $T$ has the form
$$(TJ)(x)=\min_{u\in U}\max_{v\in V}H(x,u,v,J).\xdef\tmapminimaxtwo{\lab}\eqnum\show{oneo}$$
Assuming that the termination state $t$ is reachable with probability one under all policy pairs, it can be shown that the mapping $H$ satisfies  the contraction Assumption 1.1, so results and algorithms that are similar to the ones for the preceding example apply. This reachability assumption, however, is restrictive and is not satisfied when the problem has a semicontractive character, whereby $T_{\m,\n}$ is a contraction under some policy pairs but not for others. In this case the  analysis is more complicated and requires the notion of proper and improper policies from single-player stochastic shortest path problems; see the papers [BeT91], [PaB99], [YuB13], [Yu14], and the book [Ber18]. 
\endexample

In the next section, we will view our abstract minimax problem, involving the Bellman equation \belfixedpoint,  as an optimization by a single player who minimizes against a worst-case response by an antagonistic opponent/maximizer, and we will describe the corresponding PI algorithm, which is well known both in abstract DP and Markov games. We will highlight the main weakness of this algorithm: the computational cost of the policy evaluation operation, which involves the solution of the maximizer's problem for a fixed policy of the minimizer. We will then discuss an attractive proposal by Pollatschek and Avi-Itzhak [PoA69] that overcomes this difficulty, albeit with an algorithm that requires restrictive assumptions for its validity. In Section 3, we will introduce and analyze our new algorithm, which maintains the attractive structure of the  Pollatschek and Avi-Itzhak algorithm without requiring restrictive assumptions. We will also show the validity of our algorithm in the context of a distributed asynchronous implementation, as well as in an on-line context, which involves one-state-at-a-time policy improvement, with the states generated by an underlying dynamic system or Markov chain.

\vskip-1.5pc

\section{Relations to Single-Player Abstract DP Formulations}
\mark{Relations to Single-Player Abstract DP Formulations}
\vskip-1.5pc

\pn For the single-player problem, where there is no maximizer, a contractive abstract framework for infinite horizon DP has been studied for a long time, first by Denardo [Den67] in the case of an unweighted sup-norm contraction, and then by several other authors. Denardo's paper and the subsequent book by Bertsekas and Shreve [BeS78], Chapter 4, also provide a discussion of algorithmic issues of abstract DP, including some analysis related to PI algorithms. The author's DP textbook [Ber12] and abstract DP monograph [Ber18] provide a treatment for the more general case of a weighted sup-norm contraction, and related extensions to semicontractive problems, where the Bellman operator $T_\m$ is a contraction for some policies $\m$, but not for others (e.g., stochastic shortest path problems with both proper policies that are guaranteed to reach the termination state, and improper policies that are not). The  books [Ber12] and [Ber18] also include detailed discussions of abstract DP algorithms, including various forms of PI. 

A series of joint papers on enhanced PI algorithms, by the author and Huizhen Yu [BeY10], [BeY12], [YuB13],  is particularly relevant to the present work, and served to motivate the structure of our new PI algorithms, to be given in Section 3. These papers aimed, among others, to construct (single-player) PI algorithms that were provably convergent under distributed and asynchronous implementation, with state space partitioning, and deal more effectively with improper policies in stochastic shortest path problems. In Section 3, we will discuss in some detail their connections with the PI algorithms of this paper.

In this section, we will reformulate our minimax problem of finding a fixed point of the minimax operator $T$  of Eq.\ \minimaxt\ [cf.\ the Bellman equation \belfixedpoint] as a single-player optimization problem by redefining $T$  in terms of the mapping $\ol H$ given by
$$\ol H(x,u,J)=\sup_{v\in V(x)}H(x,u,v,J),\qquad x\in X,\ u\in U(x),\ J\in B(X).\xdef\minimaxh{\lab}\eqnum\show{oneo}$$
In particular, we write $T$ as
$$(TJ)(x)=\inf_{u\in U(x)}\ol H(x,u,J),\qquad x\in X,\eqnum\show{oneo}$$
or equivalently, by introducing for each $\m\in{\cal M}$ the operator $\ol T_\m$ given by
$$(\ol T_\m J)(x)=\ol H\big(x,\m(x),J\big)=\sup_{v\in V(x)}H\big(x,\m(x),v,J\big),\qquad x\in X,\xdef\minimaxtm{\lab}\eqnum\show{oneo}$$
we  write $T$ as
$$(TJ)(x)=\inf_{\m\in {\cal M}}(\ol T_\m J)(x),\qquad x\in X.\xdef\minimaxttt{\lab}\eqnum\show{oneo}$$
Our contraction assumption implies that all the operators $\ol T_\m$, $\m\in{\cal M}$, as well as the operator $T$ are weighted sup-norm contractions from $B(X)$ to $B(X)$, with modulus $\a$.

The single-player weighted sup-norm contractive DP framework of the author's books [Ber12] and [Ber18] applies directly to the operator $T$ as defined by Eq.\ \minimaxttt. In particular, to apply this framework to a minimax problem, we start from the mapping $\ol H$ of Eq.\ \minimaxh, which defines $\ol T_\m$ via Eq.\ \minimaxtm, and then $T$, using Eq.\ \minimaxttt.

\subsubsection{PI Algorithms}

\pn In view of the preceding transformation of our minimax problem to the single-player abstract DP formalism, the PI algorithms developed for the latter apply, and in fact these algorithms have been known for a long time for the special case of finite-state Markov games, cf.\ Examples \exampleone\ and \exampletwo.

In particular, the standard form of PI generates iteratively a sequence of policies $\{\m^t\}$. The typical iteration starts with $\m^t$ and computes $\m^{t+1}$ with a minimization that involves the optimal cost function of a maximizer's abstract DP problem with the minimizer's policy fixed at $\m^t$, as follows:\footnote{\dag}{\ninepoint Policy improvement involves an optimization operation that defines the new/improved policy. Throughout this paper, and in the context of PI algorithms, we implicitly assume that this optimization can be carried out, i.e., that the optimum is attained, and write accordingly ``min" and ``max" in place of ``inf" and ``sup," respectively.}

\texshopboxnb{\pn {\bf Iteration $(t+1)$ of Abstract PI Algorithm from the Minimizer's Point of View}
\smskip
\pn Given $\m^t$,  generate $\m^{t+1}$ with a two-step process:
\nitem{(a)} {\bf Policy evaluation\/}, which computes $J_{\m^t}$ as the unique fixed point of the mapping $\ol T_{\m^t}$ given by Eq.\ \minimaxtm, i.e., 
$$J_{\m^t}=\ol T_{\m^t}J_{\m^t},\xdef\poleval{\lab}\eqnum\show{oneo}$$
or equivalently
$$J_{\m^t}(x)=\max_{v\in V(x)}H\big(x,\m^t(x),v,J_{\m^t}\big),\qquad x\in X.\xdef\polevalexplicit{\lab}\eqnum\show{oneo}$$
\nitem{(b)} {\bf Policy improvement\/}, which computes $\m^{t+1}$ as a policy that satisfies
$$\ol T_{\m^{t+1}}J_{\m^t}=TJ_{\m^t},\eqnum\show{oneo}$$}\texshopboxnt{\pn
or equivalently
$$\m^{t+1}(x)\in \arg\min_{u\in U(x)}\left(\max_{v\in V(x)}H(x,u,v,J_{\m^t})\right),\qquad x\in X.\xdef\polimproveexplicit{\lab}\eqnum\show{oneo}$$
}

There are also {\it optimistic forms of PI\/}, which starting with a function $J^0\in B(X)$, generate  a sequence of function-policy pairs $\{J^t,\m^t\}$ with the algorithm
$$\ol T_{\m^t}J^{t}=TJ^{t},\qquad J^{t+1}=\ol T_{\m^t}^{m_t}J^{t},\qquad k=0,1,\ldots,\xdef\synchropt{\lab}\eqnum\show{oneo}$$
where 
$\{m_t\}$ is a sequence of positive integers (see e.g., [Ber12], Section 2.5.5). Here the policy evaluation operation \poleval\ that finds the fixed point of the mapping $\ol T_{\m^t}$ is approximated by $m_t$ value iterations using $\ol T_{\m^t}$, and starting from $J^{t}$, as in the second equation of \synchropt. The convergence of the abstract forms of these PI algorithms has been established under the additional monotonicity assumption
$$\ol T_\m J\le \ol T_\m J'\qquad\hbox{for all } J,J'\in B(X)\hbox{ with }J\le J',\xdef\monotonicity{\lab}\eqnum\show{oneo}$$
which is typically satisfied in DP-type single-player and two-player problem formulations (function inequalities are meant to be pointwise in this paper); proofs are given in the book [Ber18], Sections 2.4 and 2.5, which provides references to earlier works.

The drawback of the preceding PI algorithms is that the policy evaluation operation of Eq.\ \poleval\ and its optimistic counterpart of Eq.\ \synchropt\ aim to find or approximate the fixed point of $\ol T_{\m^t}$, which  involves a potentially time-consuming maximization over $v\in V(x)$; cf.\ the definition \minimaxtm\ and Eq.\ \polevalexplicit. This can be seen from the fact that Eq.\ \polevalexplicit\ is Bellman's equation for a maximizer's abstract DP problem, where the minimizer is known to use the policy $\m^t$. There is a PI algorithm for finite-state Markov games, due to Pollatschek and Avi-Itzhak [PoA69], which was specifically designed to avoid the use of maximization over $v\in V(x)$ in the policy evaluation operation. We present this algorithm next, together with a predecessor PI algorithm, due to Hoffman and Karp [HoK66], which is in fact the algorithm \poleval-\polimproveexplicit\ applied to the Markov game Example \exampleone.

\subsubsection{The Hoffman-Karp, and Pollatschek and Avi-Itzhak Algorithms for Finite-State Markov Games}

\pn The PI algorithm \poleval-\polimproveexplicit\ for the special case of finite-state Markov games (cf.\ Example \exampleone), has been proposed by Hoffman and Karp [HoK66]. It takes the form
$$J_{\m^t}(x)=\max_{v\in V}H\big(x,\m^t(x),v,J_{\m^t}\big),\qquad x\in X,\xdef\polevalmarkov{\lab}\eqnum\show{oneo}$$
where $H$ is the Markov game mapping \gamemap\ (this is the policy evaluation step), followed by solving the static minimax problem
$$\min_{u\in U}\max_{v\in V}H(x,u,v,J_{\m^t}),\qquad x\in X,\xdef\polimprovemarkov{\lab}\eqnum\show{oneo}$$
and letting $\m^{t+1}$ be a policy that attains the minimum above (this is the policy improvement step). The policy improvement subproblem \polimprovemarkov\ is a matrix saddle point problem, involving the matrix 
$$A(x)+\sum_{y\in X}Q_{xy}J_{\m^t}(y),$$
[cf.\ Eq.\ \tmapminimaxone], which is easily solvable by linear programming for each $x$ (this is well-known in the theory of matrix games). 

However, the policy evaluation step \polevalmarkov\ involves the solution of the maximizer's Markov decision problem, for the fixed policy $\m^t$ of the minimizer. This can be a quite difficult problem that requires an expensive computation. The same is true for a modified version of the Hoffman-Karp algorithm proposed by van der Wal [Van76], which involves an approximate policy evaluation, based on a limited number of value iterations, as in the optimistic PI algorithm \synchropt. 
The computational difficulty of the policy evaluation phase of the Hoffman-Karp algorithm  is also shared by other PI algorithms for sequential games that have been suggested in the literature in subsequent works (e.g., Patek and Bertsekas [PaB99], and Yu [Yu14]). We refer to Akian and Gaubert [AkG01] for analysis of computational complexity issues  relating to the Hoffman-Karp algorithm.

Following the publication of the Hoffman-Karp algorithm, another PI algorithm for finite-state Markov games was proposed by Pollatschek and Avi-Itzhak [PoA69], and has attracted considerable attention because it is more computationally expedient. It generates a sequence of minimizer-maximizer policy pairs $\{\m^t,\n^t\}$ and corresponding game value functions $J_{\m^t,\n^t}(x)$, starting from each state $x$. In particular, the standard form of PI generates iteratively a sequence of policies $\{\m^t\}$. We give this algorithm in an abstract form, which parallels the PI algorithm \poleval-\polimproveexplicit. The typical iteration starts with a  pair $(\m^t,\n^t)$ and computes a pair $(\m^{t+1},\n^{t+1})$ as follows:

\texshopbox{\pn {\bf Iteration $(t+1)$ of the Pollatschek and Avi-Itzhak  PI Algorithm in Abstract Form}
\smskip
\pn Given $(\m^t,\n^t)$,  generate $(\m^{t+1},\n^{t+1})$ with a two-step process:
\nitem{(a)} {\bf Policy evaluation\/}, which computes $J_{\m^t,\n^t}$ by solving the fixed point equation
$$J_{\m^t,\n^t}(x)=H\big(x,\m^t(x),\n^t(x),J_{\m^t,\n^t}\big),\qquad x\in X.\xdef\polevalmarkovPoA{\lab}\eqnum\show{oneo}$$
\nitem{(b)} {\bf Policy improvement\/}, which computes  $(\m^{t+1},\n^{t+1})$ by solving the saddle point problem
$$\min_{u\in U}\max_{v\in V}H(x,u,v,J_{\m^t,\n^t}),\qquad x\in X.\xdef\polimproveminmax{\lab}\eqnum\show{oneo}$$
}

The Pollatschek and Avi-Itzhak algorithm  [PoA69] is the algorithm \polevalmarkovPoA-\polimproveminmax, specialized to the Markov game case of the mapping $H$ that involves the matrix 
$$A(x)+\sum_{y\in X}Q_{xy}J_{\m^t,\n^t}(y),$$
similar to the Hoffman-Karp algorithm, cf.\ Eq. \tmapminimaxone. A key observation is that the policy evaluation operation \polevalmarkovPoA\  is computationally comparable to policy evaluation in a single-player Markov decision problem, i.e., solving a linear system of equations. In particular, it does not involve solution of the Markov decision problem of the maximizer like the Hoffman-Karp PI algorithm [cf.\ Eq.\ \polevalmarkov], or its approximate solution by multiple value iterations, as in the van der Wal optimistic version \synchropt\ for Markov games.

Computational studies have shown that the Pollatschek and Avi-Itzhak algorithm converges much faster than its competitors, {\it when it converges} (see Breton et al.\ [BFH86], and also Filar  and Tolwinski [FiT91], who proposed a modification of the algorithm). Moreover, the number of iterations required for convergence is fairly small. This is consistent with an interpretation given by Pollatschek and Avi-Itzhak in their paper [PoA69], where they have shown that their algorithm coincides with a form of Newton's method for solving the fixed point/Bellman equation $J=TJ$ (see Fig.\ 2.1).\footnote{\dag}{\ninepoint Newton's method for solving a general fixed point problem of the form $z=F(z)$, where $z$ is an  $n$-dimensional vector, operates as follows: At the current iterate $z_k$, we linearize $F$ and find the solution $z_{k+1}$ of the corresponding linear fixed point problem, obtained using a first order Taylor expansion:
$$z_{k+1}=F(z_k)+{\partial F(z_k)\over \partial z}(z_{k+1}-z_k),$$
where ${\partial F(z_k)/\partial z}$ is the $n\times n$ Jacobian matrix of $F$ evaluated at the $n$-dimensional vector $z_k$.
The most commonly given convergence rate property of Newton's method is {\it quadratic convergence\/}. It states that near the solution $z^*$, we have 
$$\|z_{k+1}-z^*\|=O\big(\|z_{k}-z^*\|^2\big),$$
 where $\|\cdot\|$ is the Euclidean norm, and holds assuming the Jacobian matrix exists and is Lipschitz continuous (see [Ber16], Section 1.4). Qualitatively similar results hold under other assumptions. In particular a superlinear convergence statement (suitably modified to account for lack of differentiability of $F$) can be proved for the case where $F(z)$ has components that are either monotonically increasing or monotonically decreasing, and either concave or convex. In the case of the Pollatschek and Avi-Itzhak algorithm, the main difficulty is that the concavity/convexity condition is violated; see Fig.\ 2.1.} The close connection of PI with Newton's method is well-known in control theory and operations research, through several works, including Kleinman [Kle68] for linear-quadratic optimal control problems, and Puterman and Brumelle [PuB78], [PuB78] for more abstract settings. Its significance in reinforcement learning contexts has been discussed at length in the author's recent book [Ber20].

{
\midinsert

\centerline{\hskip0pc\epsfxsize = 5.4in \epsfbox{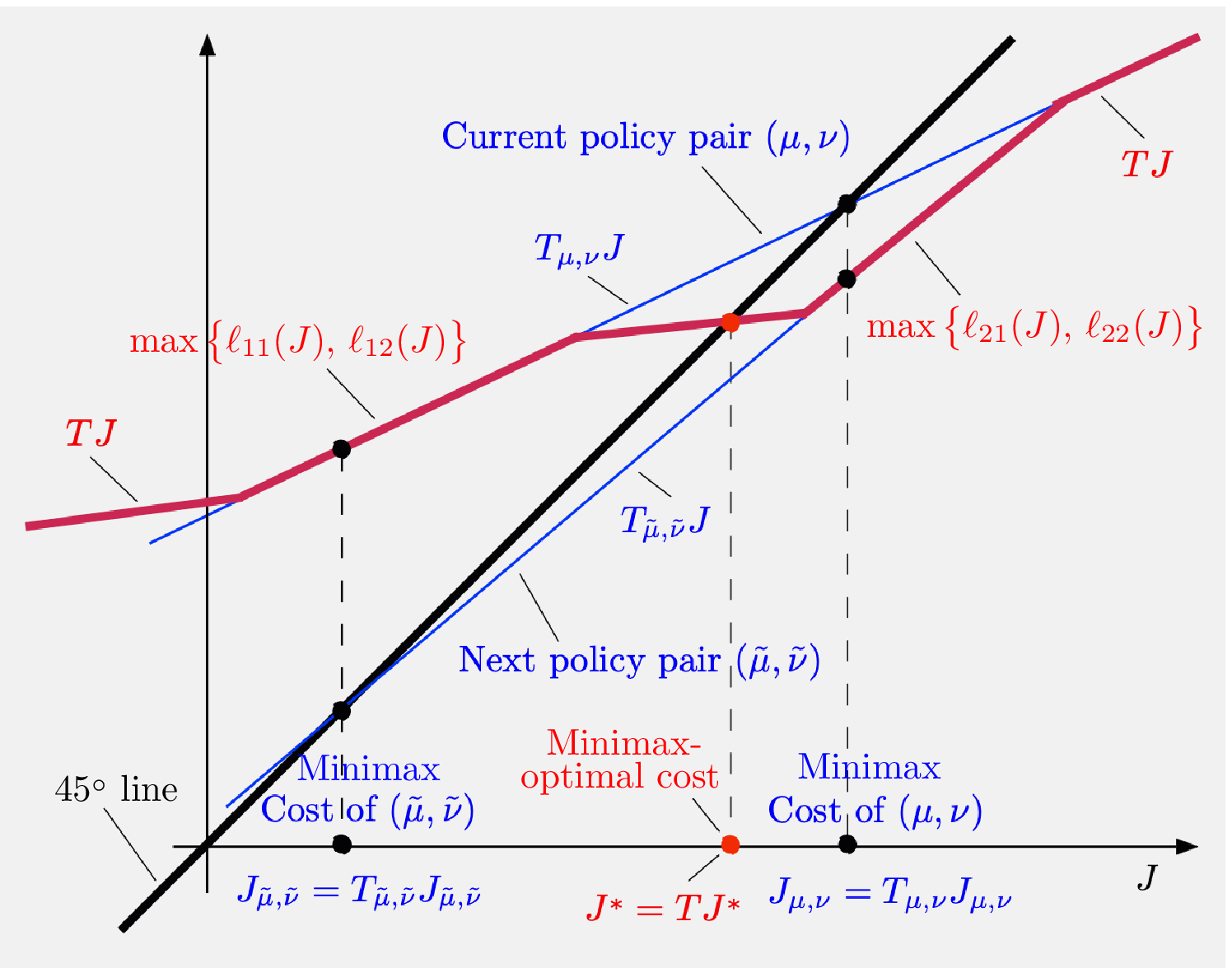}}
\vskip-1pc
\fig{0pc}{2.1.} {Schematic illustration of the abstract minimax  PI algorithm \polevalmarkovPoA-\polimproveminmax\ in the case of a minimax problem involving a single state, in addition to a termination state $t$; cf.\ Example \exampletwo. We have $J^*(t)=0$ and $(TJ)(t)=0$ for all $J$ with $J(t)=0$, so that the operator $T$ can be graphically represented in just one dimension (denoted by $J$) that corresponds to the nontermination state. This makes it easy to visualize $T$ and geometrically interpret why Newton's method does not converge. 
\old{In particular, for the case of two possible controls for each of the minimizer and maximizer, the operator $T$ has the form
$$TJ=\min_{(u_1,\,u_2)\in U}\max_{(v_1,\,v_2)\in V}\big(u_1\,\, u_2\big)\left(\left[\matrix{a_{11}&a_{12}\cr a_{21}&a_{22}\cr}\right]+\left[\matrix{q_{11}&q_{12}\cr q_{21}&q_{22}\cr}\right]J\right)\pmatrix{v_1\cr v_2\cr};$$
where the $2\times 2$ matrices above correspond to the ones of Eqs.\ \tmapminimaxone-\tmapminimaxtwo.}
Because the operator $T$ may be neither convex nor concave for a minimax problem, the algorithm may cycle between pairs $(\m,\n)$ and $(\tl\m,\tl \n)$, as shown in the figure. By contrast in a (single-player) finite-state Markovian decision problem, $T$ has piecewise linear and concave components, and the PI algorithm converges in a finite number of iterations.  
The figure illustrates an operator $T$ of the form
$$TJ=\min\Big\{\max\big\{\ell_{11}(J),\,\ell_{12}(J)\big\},\,\max\big\{\ell_{21}(J),\,\ell_{22}(J)\big\}\Big\},$$
where $\ell_{ij}(J)$, are linear functions of $J$, corresponding to the choices $i=1,2$ of the minimizer and $j=1,2$ of the maximizer. Thus $TJ$ is the minimum of the convex functions 
$$\max\big\{\ell_{11}(J),\,\ell_{12}(J)\big\}\qquad  \hbox{and}\qquad \max\big\{\ell_{21}(J),\,\ell_{22}(J)\big\},$$
as shown in the figure. Newton's method linearizes $TJ$ at the current iterate [i.e., replaces $TJ$ with one of the four linear functions $\ell_{ij}(J)$, $i=1,2$, $j=1,2$ (the one attaining the min-max at the current iterate)] and solves the corresponding linear fixed point problem to obtain the next iterate.
}\endinsert
}

Unfortunately, however, the Pollatschek and Avi-Itzhak algorithm is valid  only under restrictive assumptions (given in their paper [PoA69]). The difficulty is that Newton's method applied to the Bellman equation $J=TJ$ need not be globally convergent when the operator $T$ corresponds to a minimax problem. This is illustrated in Fig.\ 2.1, which also illustrates why Newton's method (equivalently, the PI algorithm)  is globally convergent in the case of a single-player finite-state Markov decision problem, as is well known. In this case each component $(TJ)(x)$ of the function $TJ$ is concave and piecewise linear, thereby guaranteeing the finite termination of the PI algorithm. This is not true in the case of finite-state minimax problems and Markov games. The difficulty is that {\it the  functions $(TJ)(x)$ may be neither convex nor concave in $J$\/}, even though they are piecewise linear and have the monotonicity property \monotonicity\ (cf.\ Fig.\ 2.1).
In fact a two-state example where the Pollatschek and Avi-Itzhak algorithm does not converge to $\jstar$ was given by  van der Wal [Van76]. This example involves a single state in addition to a termination state, and the algorithm oscillates similar to  Fig.\ 2.1.  Note that the Hoffman-Karp algorithm does not admit an interpretation as Newton's method, and is not subject to the convergence difficulties of the Pollatschek and Avi-Itzhak algorithm.

\vskip-1.5pc

\section{A New PI Algorithm for Abstract Minimax DP problems}
\mark{A New PI Algorithm for Abstract Minimax DP problems}
\vskip-1.5pc

\pn In this section, we will introduce modifications to the Pollatschek and Avi-Itzhak algorithm, and its abstract version \polevalmarkovPoA-\polimproveminmax, given in the preceding section, with the aim to enhance its convergence properties, while maintaining its favorable structure.  These modifications will apply to a general minimax problem of finding a fixed point of a suitable contractive operator, and offer the additional benefit that they allow asynchronous, distributed, and on-line implementations.

Our PI algorithm is motivated by a line of analysis and corresponding algorithms  introduced by Bertsekas and Yu [BeY10], [BeY12] for discounted infinite horizon DP problems, and by Yu and Bertsekas [YuB13] for stochastic shortest path problems (with both proper and improper policies). These algorithms were also presented in  general abstract form in several of the author's books [Ber12], Section 2.6.3, [Ber18], Sections 2.6.3 and 3.6.2, and [Ber20], Chapter 5. The PI algorithm of this section uses a similar abstract formulation, but replaces the single mapping that is minimized in these works with two mappings, one of which is minimized while the other is maximized. Mathematically, the difficulty of the Pollatschek and Avi-Itzhak algorithm is that the policies $(\m^{t+1},\n^{t+1})$ obtained from the policy improvement/static game \polimproveminmax\ are not ``improved" in a clear sense, such as 
$$J_{\m^{t+1},\n^{t+1}}(x)\le J_{\m^{t},\n^{t}}(x),\qquad \hbox{for all }x\in X,$$
as they are in the case of single-player DP, where a policy improvement property is central in the standard convergence proof of single-player PI. Our algorithm, however, does not rely on policy improvement, but rather derives its validity from a {\it uniform contraction property of an underlying operator\/}, to be given in Section 4 (cf.\ Prop.\ 4.2). In fact, {\it our algorithm does not require the monotonicity assumption \monotonicity\ for its convergence\/}, and thus it can be used in minimax problems that are beyond the scope of DP.

As an aid to understanding intuitively the abstract framework of this section, we note that it is patterned after a multistage process, whereby at each stage, the following sequence of events is envisioned (cf.\ Fig.\ 3.1):
\nitem{(1)} We start at some state $x_1$ from a space $X_1$.
\nitem{(2)} The minimizer, knowing $x_1$, chooses a control $u\in U(x_1)$. Then a new state $x_2$ from a space $X_2$ is generated as a function of $(x_1,u)$. (It is possible that $X_1=X_2$, but for greater generality, we do not assume so. Also the transition from $x_1$ to $x_2$ may involve a random disturbance; see the subsequent Example 3.4.)
\nitem{(3)} The maximizer, knowing $x_2$, chooses a control $v\in V(x_2)$. Then a new state $\ol x_1\in X_1$ is generated.
\nitem{(4)} The next stage is started at $\ol x_1$ and the process is repeated.
\smskip
\pn If we start with $x_1\in X_1$, this sequence of events corresponds to finding the optimal minimizer policy against a worst case choice of the maximizer, and the corresponding min-max value is denoted by $\jstar_1(x_1)$. Symmetrically, if we start with $x_2\in X_2$, this sequence of events corresponds to finding the optimal maximizer policy against a worst case choice of the minimizer, and the corresponding max-min value is denoted by $\jstar_2(x_2)$.

{
\topinsert
\centerline{\hskip0pc\epsfxsize = 4.0in \epsfbox{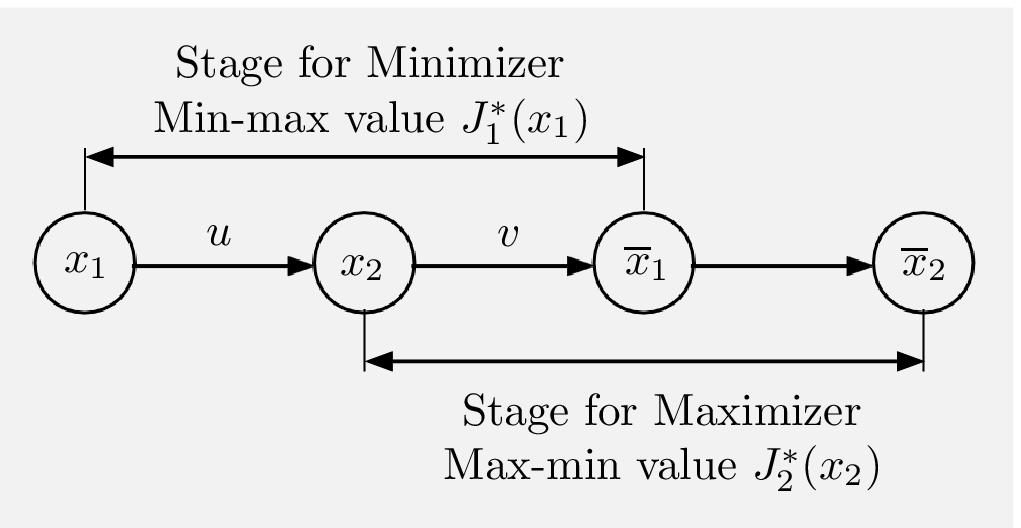}}
\vskip-1pc
\fig{0pc}{3.1.} {Schematic illustration of the sequence of events at each stage of the minimax problem. We start at  $x_1\in X_1$.
The minimizer chooses a control $u\in U(x_1)$, a new state $x_2\in X_2$ is generated,
the maximizer chooses a $v\in V(x_2)$, and a new state $\ol x_1\in X_1$ is generated, etc. 
If the stage begins at $x_2$ rather than $x_1$, this corresponds to the max-min problem. The corresponding min-max and max-min values are $J^*_1(x_1)$ and $J^*_2(x_2)$, respectively.}
\endinsert
}

This type of framework can be viewed within the context of the theory of zero-sum games in extensive form, a methodology with a long history [Kuh53]. Games in extensive form involve sequential/alternating choices by the players with knowledge of prior choices. By contrast, for games in simultaneous form, such as the Markov games of the preceding section, the  players make their choices without being sure of the other player's choices.

\vskip-0.5pc

\subsubsection{Fixed Point Formulation}

\pn We consider the space of bounded functions of $x_1\in X_1$, denoted by $B(X_1)$, and the space of bounded functions of $x_2\in X_2$, denoted by $B(X_2)$, with respect to the norms
$\|J_1\|_1$ and $\|J_2\|_2$ defined by
$$\|J_1\|_1=\sup_{x_1\in X_1}{\big|J_1(x_1)\big|\over \xi_1(x_1)},\qquad \|J_2\|_2=\sup_{x_2\in X_2}{\big|J_2(x_2)\big|\over \xi_2(x_2)},\xdef\weightednorms{\lab}\eqnum\show{oneo}$$
where $\xi_1$ and $\xi_2$ are positive weighting functions, respectively. 
We also consider the space $B(X_1)\times B(X_2)$ with the norm
$$\big\|(J_1,J_2)\big\|=\max\big\{\|J_1\|_1,\,\|J_2\|_2\big\}.\xdef\supnormtwo{\lab}\eqnum\show{oneo}$$

We will be interested in finding a pair of functions $(\jstar_1,\jstar_2)$ that are the fixed point of mappings 
$$H_1:X_1\times U\times B(X_2)\mapsto B(X_1),\qquad H_2:X_2\times V\times B(X_1)\mapsto B(X_2),$$
in the following sense: for all $x_1\in X_1$ and $x_2\in X_2$,  
$$\jstar_1(x_1)=\inf_{u\in U(x_1)}H_1(x_1,u,\jstar_2),\qquad \jstar_2(x_2)=\sup_{v\in V(x_2)}H_2(x_2,v,\jstar_1).\xdef\doublefix{\lab}\eqnum\show{oneo}$$
These two equations form an abstract version of Bellman's equation for the infinite horizon sequential min-max problem described by the sequence of events (1)-(4) given earlier.  
We will assume later (see Section 4) that $H_1$ and $H_2$ have a contraction property like Assumption 1.1, which will guarantee that $(\jstar_1,\jstar_2)$ is the unique fixed point within $B(X_1)\times B(X_2)$.

Note that the fixed point problem \doublefix\ involves both min-max and max-min values, without assuming that they are equal. By contrast the algorithms of Section 2 aim to compute only the min-max value. In the case of a Markov game (cf.\ Examples \exampleone\ and \exampletwo),  the min-max value is equal to the max-min value, but in general min-max may not be equal to max-min, and the algorithms of Section 2 will only find min-max explicitly. We will next provide an example to interpret $\jstar_1$ and $\jstar_2$ as the min-max and max-min value functions of a sequential infinite horizon problem involving the sequence of events (1)-(4) given earlier.

\xdef\exampleminimaxcontroltwo{\exampl}\examplnum\show{myexample}

\beginexample{\exampleminimaxcontroltwo\ (Discounted Minimax Control - Explicit Separation of the Two Players)}\pn In this formulation of a discounted minimax control problem, the states of the minimizer and the maximizer, respectively, at time $k$ are denoted by  
 $x_{1,k}\in X_1$ and $x_{2,k}\in X_2$, and they evolve according to
  $$x_{2,k+1}=f_1(x_{1,k},u_k),\qquad x_{1,k+1}=f_2(x_{2,k+1},v_k),\qquad k=0,1,\ldots.\xdef\minmaxone{\lab}\eqnum\show{oneo}$$
The mappings $H_1$ and $H_2$ are given by
$$H_1(x_1,u,J_2)=g_1(x_1,u)+\a J_2\big(f_1(x_1,u)\big),\qquad H_2(x_2,v,J_1)=g_2(x_2,v)+\a J_1\big(f_2(x_2,v)\big),\xdef\minmaxthree{\lab}\eqnum\show{oneo}$$
where $g_1$ and $g_2$ are stage cost functions for the minimizer and the maximizer, respectively.
The corresponding fixed point problem of Eq.\ \doublefix\ has the form
$$J_1^*(x_1)=\inf_{u\in U(x_1)}\Big[g_1(x_1,u)+\a J_2^*\big(f_1(x_1,u)\big)\Big],\qquad J_2^*(x_2)=\sup_{v\in V(x_2)}\Big[g_2(x_2,v)+\a J_1^*\big(f_2(x_2,v)\big)\Big].\xdef\minmaxfive{\lab}\eqnum\show{oneo}$$
\endexample

\xdef\examplefour{\exampl}\examplnum\show{myexample}

\beginexample{\examplefour\ (Markov Games)}We will show that the discounted Markov game of Example \exampleone\ can be reformulated within our fixed point framework of Eq.\ \doublefix\ by letting $X_1=X$, $X_2=X\times U$, and by redefining the minimizer's control to be a probability distribution $(u_1,\ldots,u_n)$, and the maximizer's control to be one of the $m$ possible choices $j=1,\ldots,m$. 

To introduce into our problem formulation an appropriate contraction structure that we will need in the next section, we use a scaling parameter $\b$ such that 
$$\b>1,\qquad \a\b<1.\xdef\scalingpar{\lab}\eqnum\show{oneo}$$
 The idea behind the use of the scaling parameter $\b$ is to introduce discounting into the stages of both the minimizer and the maximizer. We consider functions  $J_1^*(x)$ and $J_2^*(x,u)$  that solve the equations
$$J_1^*(x)={1\over \b}\min_{u\in U}J_2^*(x,u),\xdef\markovbelone{\lab}\eqnum\show{oneo}$$
$$J_2^*(x,u)=\max\left\{ u'\left(A(x)+\a\b \sum_{y\in X}Q_{xy}J_1^*(y)\right)(1),\ldots,u'\left(A(x)+\a\b \sum_{y\in X}Q_{xy}J_1^*(y)\right)(m)\right\},\xdef\markovbeltwo{\lab}\eqnum\show{oneo}$$
where
$$\left(A(x)+\a\b \sum_{y\in X}Q_{xy}J_1^*(y)\right)(j),\qquad j=1,\ldots,m,\xdef\markovcolumn{\lab}\eqnum\show{oneo}$$
denotes the $j$th column of the matrix
$$A(x)+\a\b\sum_{y\in X}Q_{xy}J_1^*(y).\xdef\markovmatrix{\lab}\eqnum\show{oneo}$$
 It can be seen from these equations that 
$$J_2^*(x,u)=\max_{v\in V}\,u'\left(A(x)+\a\b \sum_{y\in X}Q_{xy}J_1^*(y)\right)v,\xdef\markovbelthree{\lab}\eqnum\show{oneo}$$
since the maximization over $v\in V$ above is equivalent to the maximization over the $m$ alternatives in Eq.\ \markovbeltwo, which correspond to the extreme points of the unit simplex $V$. Thus from Eqs.\ \markovbelone\ and \markovbelthree, it follows that the function $\b J_1^*$ satisfies
$$(\b J_1^*)(x)=\min_{u\in U}\max_{v\in V}\,u'\left(A(x)+\a\sum_{y\in X}Q_{xy}(\b J_1^*)(y)\right)v,$$
so it coincides with the vector of equilibrium values $J^*$ of the Markov game formulation of Example \exampleone\ [cf.\ Eq.\ \gamemap-\bellmanmarkov].

Note that $J_2^*(x,\cdot)$ is a piecewise linear function of $u$ with at most $m$ pieces, defined by the columns \markovcolumn. Thus the fixed point $(J_1^*,J_2^*)$ can be stored and be computed as a finite set of numbers: the real numbers $J_1^*(x)$, $x\in X$, which can also be used to compute the $n\times m$ matrices $A(x)+\a\b\sum_{y\in X}Q_{xy}J_1^*(y)$, $x\in X$, whose columns define $J^*_2(x,u)$, cf.\ Eq.\ \markovbeltwo. 

We finally observe that the two equations \markovbelone\ and \markovbelthree\ can be written in the form \doublefix, with $x_1=x$, $x_2=(x,u)$, and $H_1$, $H_2$  defined by
$$H_1(x,u,J_2)={1\over \b}J_2(x,u),\qquad H_2(x,u,v,J_1)=u'\left(A(x)+\a\b \sum_{y\in X}Q_{xy}J_1^*(y)\right)v.$$
\endexample

An important area of application of our two-player framework is control under set-membership uncertainty within a game-against-nature formulation, whereby nature is modeled as an antagonistic opponent choosing $v\in V(x_2)$. Here only the min-max value is of practical interest, but our subsequent PI methodology will find the max-min value as well. We provide two examples of this type of formulation.

\xdef\exampleminimaxcontrol{\exampl}\examplnum\show{myexample}

\beginexample{\exampleminimaxcontrol\ (Discounted Minimax Control Over an Infinite Horizon)}\pn Consider a dynamic system whose state evolves at each time $k$ according to a discrete time equation of the form
$$x_{k+1}=f(x_k,u_k,v_k),\qquad k=0,1,\ldots,\xdef\syseq{\lab}\eqnum\show{oneo}$$
where $x_k$ is the state, $u_k$ is the control to be selected from some given set $U(x_k)$ (with perfect knowledge of $x_k$), and $v_k$ is a disturbance that is selected by an antagonistic nature from a set $V(x_k,u_k)$ [with perfect knowledge of $(x_k,u_k)$]. A cost $g(x_k,u_k,v_k)$ is incurred at time $k$, it is accumulated over an infinite horizon, and it is discounted by $\a\in(0,1)$.  The Bellman equation for this problem is 
$$J^*(x)=\inf_{u\in U(x)}\sup_{v\in V(x,u)}\Big[g(x,u,v)+\a J^*\big(f(x,u,v)\big)\Big],\xdef\minimaxbellman{\lab}\eqnum\show{oneo}$$
and the optimal cost function $J^*$ is the unique fixed point of this equation, assuming that the cost per stage $g$ is a bounded function.

To reformulate this problem into the fixed point format \doublefix, we identify the minimizer's state $x_1$ with the state $x$ of the system \syseq, and the maximizer's state $x_2$ with the state-control pair $(x,u)$. We also introduce a scaling parameter $\b$ that satisfies Eq.\ \scalingpar. We define $H_1$ and $H_2$ as follows:
$$H_1(x,u,J_2)\hbox{ maps $(x,u,J_2)$ to the real value }{1\over \b}J_2(x,u),$$
$$H_2(x,u,v,J_1)\hbox{ maps $(x,u,v,J_1)$  to the real value }g(x,u,v)+\a\b J_1\big(f(x,u,v)\big).$$
Then the resulting fixed point problem \doublefix\ takes the form
$$(\b J_1^*)(x)=\inf_{u\in U(x)}J_2^*(x,u),\qquad J_2^*(x,u)=\sup_{v\in V(x,u)}\Big[g(x,u,v)+\a(\b J_1^*)\big(f(x,u,v)\big)\Big],$$
which is equivalent to the Bellman equation \minimaxbellman\ with $J^*=\b J^*_1$.
\endexample

\xdef\exampleminimaxcontrolpartial{\exampl}\examplnum\show{myexample}

\beginexample{\exampleminimaxcontrolpartial\ (Discounted Minimax Control with Partially Stochastic Disturbances)}\pn Consider a dynamic system such as the one of Eq.\ \syseq\ in the preceding example, except that there is an additional stochastic disturbance $w$ with known conditional probability distribution given $(x,u,v)$. Thus the state evolves at each time $k$ according to
$$x_{k+1}=f(x_k,u_k,v_k,w_k),\qquad k=0,1,\ldots,\xdef\syseq{\lab}\eqnum\show{oneo}$$
and the cost per stage is $g(x_k,u_k,v_k,w_k)$.  The Bellman equation now is 
$$J^*(x)=\inf_{u\in U(x)}\sup_{v\in V(x,u)}E_w\Big\{g(x,u,v,w)+\a J^*\big(f(x,u,v,w)\big)\ \big|\ x,u,v\Big\},\xdef\minimaxbellmanst{\lab}\eqnum\show{oneo}$$
and $J^*$ is the unique fixed point of this equation, assuming that $g$ is a bounded function.

Similar to Example \exampleminimaxcontrol, we let the minimizer's state be $x$, and the maximizer's state be $(x,u),$ we introduce a scaling parameter $\b$ that satisfies Eq.\ \scalingpar, and we define $H_1$ and $H_2$ as follows:
$$H_1(x,u,J_2)\hbox{ maps $(x,u,J_2)$ to the real value }{1\over \b}J_2(x,u),$$
$$H_2(x,u,v,J_1)\hbox{ maps $(x,u,v,J_1)$  to the real value }E_w\Big\{g(x,u,v,w)+\a\b J_1\big(f(x,u,v,w)\big)\ \big|\ x,u,v\Big\}.$$
The resulting fixed point problem \doublefix\ takes the form
$$(\b J_1^*)(x)=\inf_{u\in U(x)}J_2^*(x,u),\qquad J_2^*(x,u)=\sup_{v\in V(x,u)}E_w\Big\{g(x,u,v,w)+\a(\b J_1^*)\big(f(x,u,v,w)\big)\ \big|\ x,u,v\Big\}.$$
which is equivalent to the Bellman equation \minimaxbellmanst\ with $J^*=\b J^*_1$.
\endexample

Other examples of application of our abstract game framework \doublefix\ include two-player versions of multiplicative and exponential cost problems. One-player cases of these problems have a long tradition in DP; see e.g., Jacobson [Jac73], Denardo and Rothblum [DeR79], Whittle [Whi81], Rothblum [Rot84], Patek [Pat01]. 
Abstract versions of these problems  come under the general framework of {\it affine monotonic problems\/}, for which we refer to the author's book [Ber18a] (Section 3.5.2) and paper [Ber19a] for further discussion.
Two-player versions of affine monotonic problems involve a state space $X=\{1,\ldots,n\}$, and the mapping 
$$H(x,u,v,J)=g(x,u,v)+\sum_{y=1}^nA_{xy}(u,v)J(y),\qquad x=1,\ldots,n,$$
where $g$ and $A_{xy}$ satisfy
$$g(x,u,v)\ge0,\qquad A_{xy}(u,v)\ge0,\qquad \hbox{for all } x,y=1,\ldots,n,\ u\in U(x),\ v\in V(x).$$
Our PI algorithms can be suitably adapted to address these problems, along the lines of the preceding examples. Of course, the corresponding convergence analysis may pose special challenges, depending on whether our assumptions of the next section are satisfied.\old{
We will later discuss alternative forms of discounted minimax control problems, where there is no explicit separation of the two players. 
Another example of major interest is the classical discounted Markov game of Example \exampleone.  We will postpone for later its reformulation into our abstract game framework of Eq.\ \doublefix, after we introduce our PI algorithm.
}

\subsubsection{``Naive" PI Algorithms}
\pn A PI algorithm for the fixed point problem \doublefix, which is patterned after the Pollatschek and Avi-Itzhak algorithm, generates a sequence of policy pairs $\{\m^t,\n^t\}\subset {\cal M}\times{\cal N}$ and corresponding sequence of cost function pairs $\{J_{1,\m^t,\n^t},J_{2,\m^t,\n^t}\}\subset B(X_1)\times B(X_2)$. We use the term ``naive" to indicate that the algorithm does not address adequately the convergence issue of the underlying Newton's method.\footnote{\dag}{\ninepoint  We do not mean the term in a pejorative sense. In fact the Pollatschek and Avi-Itzhak paper [PoA69] embodies original ideas, includes sophisticated and insightful analysis, and has stimulated considerable followup work.}   Given $\{\m^t,\n^t\}$ it generates  $\{\m^{t+1},\n^{t+1}\}$ with a two-step process as follows:
\nitem{(a)} {\bf Policy evaluation\/}, which computes corresponding functions $\{J_{1,\m^t,\n^t,J_2^t},J_{2,\m^t,\n^t}\}$ by solving the fixed point equations
$$J_{1,\m^t,\n^t}(x_1)=H_1\big(x_1,\m^t(x),J_{2,\m^t,\n^t}\big),\qquad x_1\in X_1,\xdef\polevalmarkovPoAone{\lab}\eqnum\show{oneo}$$
$$J_{2,\m^t,\n^t}(x_2)=H_2\big(x_2,\n^t(x),J_{1,\m^t,\n^t}\big),\qquad x_2\in X_2.\xdef\polevalmarkovPoAtwo{\lab}\eqnum\show{oneo}$$
\nitem{(b)} {\bf Policy improvement\/}, which computes  $(\m^{t+1},\n^{t+1})$ with the minimizations
$$\m^{t+1}(x_1)\in\arg\min_{u\in U(x_1)} H_1\big(x_1,u,J_{2,\m^t,\n^t}\big),\qquad x_1\in X_1,\xdef\polimprPoAone{\lab}\eqnum\show{oneo}$$
$$\n^{t+1}(x_2)\in\arg\max_{v\in V(x_2)} H_2\big(x_2,v,J_{1,\m^t,\n^t}\big),\qquad x_2\in X_2.\xdef\polimprPoAtwo{\lab}\eqnum\show{oneo}$$
\smskip

This algorithm resembles the abstract version of the Pollatschek and Avi-Itzhak algorithm \polevalmarkovPoA-\polimproveminmax\ in that it involves simple policy evaluations, which do not require the solution of a multistage DP problem for either the minimizer or the maximizer. Unfortunately, however, the algorithm \polevalmarkovPoAone-\polimprPoAtwo\ cannot be proved to be convergent, as it does not deal effectively with the oscillatory behavior illustrated in Fig.\ 2.1.  

An optimistic version of the PI algorithm \polevalmarkovPoAone-\polimprPoAtwo\ evaluates the fixed point pair 
$(J_{1,\m^t,\n^t},J_{2,\m^t,\n^t})$ approximately, by using some number, say $\bar k\ge1$, of value iterations. It has the form
$$J_{1,k+1}(x_1)=H_1\big(x_1,\m^t(x_1),J_{2,k}\big),\qquad x_1\in X_1,\qquad k=0,1,\ldots,\bar k-1,\xdef\polevalPoAoptone{\lab}\eqnum\show{oneo}$$
$$J_{2,k+1}(x_2)=H_2\big(x_2,\n^t(x_2),J_{1,k}\big),\qquad x_2\in X_2,\qquad k=0,1,\ldots,\bar k-1,\xdef\polevalPoAopttwo{\lab}\eqnum\show{oneo}$$
starting from an initial approximation $(J_{1,0},J_{2,0})$, instead of solving the fixed point equations \polevalmarkovPoAone-\polevalmarkovPoAtwo. As $\bar k$ (i.e., the number of value iterations used for policy evaluation) increases, the pair $(J_{1,\bar k},J_{2,\bar k})$ converges to $(J_{1,\m^t,\n^t},J_{2,\m^t,\n^t})$, and the optimistic and nonoptimistic policy evaluations coincide in the limit (under suitable contraction assumptions to be introduced in the next section). Still the PI algorithm that uses this optimistic policy evaluation, followed by a policy improvement operation similar to Eqs.\ \polimprPoAone-\polimprPoAtwo, i.e.,
$$\m^{t+1}(x_1)\in\arg\min_{u\in U(x_1)} H_1\big(x_1,u,J_{2,\bar k+1}\big),\qquad x_1\in X_1,\xdef\polimprPoAonefinal{\lab}\eqnum\show{oneo}$$
$$\n^{t+1}(x_2)\in\arg\max_{v\in V(x_2)} H_2\big(x_2,v,J_{1,\bar k+1}\big),\qquad x_2\in X_2,\xdef\polimprPoAtwofinal{\lab}\eqnum\show{oneo}$$
cannot be proved convergent and is subject to oscillatory behavior. 
However, this optimistic algorithm can be made convergent through modifications that we describe next.

\subsubsection{Our Distributed Optimistic Abstract PI Algorithm}

\pn Our PI algorithm for finding the solution $(\jstar_1,\jstar_2)$ of the Bellman equation \doublefix\ has structural similarity with the ``naive" PI algorithm that uses optimistic policy evaluations of the form \polevalPoAoptone-\polevalPoAopttwo\ and policy improvements of the form \polimprPoAonefinal-\polimprPoAtwofinal. It differs from the PI algorithms of the preceding section, such as the Hoffman-Karp and van der Wal algorithms, in two ways:

\nitem{(a)} {\it It treats symmetrically the minimizer and the maximizer\/}, in that it aims to find both the min-max and the max-min cost functions, which are $\jstar_1$ and $\jstar_2$, respectively, and it ignores the possibility that we may have $\jstar_1=\jstar_2$.

\nitem{(b)} {\it It separates the policy evaluations and policy improvements of the minimizer and the maximizer, in asynchronous fashion\/}. In particular, in the algorithm that we will present shortly, each iteration will consist of only one of four operations: 
(1) an approximate policy evaluation (consisting of a single value iteration) by the minimizer, (2) a policy improvement by the minimizer, (3) an approximate policy evaluation (consisting of a single value iteration) by the maximizer, (4) a policy improvement by the maximizer. 
\smskip
The order and frequency by which these four operations are performed does not affect the convergence of the algorithm, as long as all of these operations are performed infinitely often. Thus the algorithm is well suited for distributed implementation. Moreover, by executing the policy evaluation steps (1) and (3) much more frequently than the policy improvement operations (2) and (3), we obtain an algorithm involving nearly exact policy evaluation. 

Our algorithm generates {\it two} sequences of function pairs, and a sequence of policy pairs:
$$\{J_1^t,J_2^t\}\subset B(X_1)\times B(X_2),\qquad  \{V_1^t,V_2^t\}\subset B(X_1)\times B(X_2),\qquad  \{\m^t,\n^t\}\subset {\cal M}\times{\cal N}.$$
The algorithm involves pointwise minimization and maximization operations on pairs of functions, which we treat notationally as follows: For any pair of functions $(V,J)$ from within $B(X_1)$ or $B(X_2)$, we denote by $\min[V,J]$ and by $\max[V,J]$ the functions defined on $B(X_1)$ or $B(X_2)$, respectively, that take values 
$$\min[V,J](x)=\min\big\{V(x),J(x)\big\},\qquad \max[V,J](x)=\max\big\{V(x),J(x)\big\},$$
for every $x$ in $X_1$ or $X_2$, respectively.

At iteration $t$, our algorithm starts with 
$$J_1^t,V_1^t,J_2^t,V_2^t,\m^t,\n^t,$$
 and generates 
 $$J_1^{t+1},V_1^{t+1},J_2^{t+1},V_2^{t+1},\m^{t+1},\n^{t+1},$$
  by executing {\it one} of the following four operations.\footnote{\dag}{\ninepoint  The choice of operation is arbitrary at iteration $t$, as long as each type of operation is executed for infinitely many $t$. It can be extended by introducing ``communication delays," and state space partitioning, whereby the operations are carried out in just a subset of the corresponding state space. This is a type of asynchronous operation that was also used in the earlier works [BeY10], [BeY12], [YuB13]. It is supported by an asynchronous convergence analysis originated in the author's papers [Ber82], [Ber83]; see also the book [BeT89], and the book [Ber12], Section 2.6. This asynchronous convergence analysis applies because the mapping underlying our algorithm is a contraction with respect to a sup-norm (rather than some other norm such as an $L_2$ norm).}

\texshopboxnb{\pn {\bf Iteration $(t+1)$ of  Distributed Optimistic Abstract PI Algorithm}
\smskip
\pn Given $(J_1^t,V_1^t,J_2^t,V_2^t,\m^t,\n^t),$ do one of the following four operations (a)-(d):
\nitem{(a)} {\bf Single value iteration for policy evaluation of the minimizer}: For all $x_1\in X_1$, set
$$J_1^{t+1}(x_1)=H_1\big(x_1,\m^t(x_1),\max[V_2^t,J_2^t]\big),\xdef\polevalmin{\lab}\eqnum\show{oneo}$$
and leave $J_2^t,V_1^t,V_2^t,\m^t,\n^t$ unchanged, i.e., the corresponding $(t+1)$-iterates are set to the $t$-iterates:
$J_2^{t+1}=J_2^t$,\ $V_1^{t+1}=V_1^t$, $V_2^{t+1}=V_2^t$, $\m^{t+1}=\m^t$, $\n^{t+1}=\n^t.$
\nitem{(b)} {\bf Policy improvement for the minimizer}: For all $x_1\in X_1$, set
$$J_1^{t+1}(x_1)=V_1^{t+1}(x_1)=\min_{u\in U(x_1)}H_1\big(x_1,u,\max[V_2^t,J_2^t]\big),\xdef\polimprmin{\lab}\eqnum\show{oneo}$$
set $\m^{t+1}(x_1)$ to a control $u\in U(x_1)$ that attains the above minimum,
and leave $J_2^t,V_2^t,\n^t$ unchanged.
\nitem{(c)} {\bf Single value iteration for policy evaluation of the maximizer}: For all $x_2\in X_2$, set
$$J_2^{t+1}(x_2)=H_2\big(x_2,\n^t(x_2),\min[V_1^t,J_1^t]\big),\xdef\polevalmax{\lab}\eqnum\show{oneo}$$
and leave $J_1^t,V_1^t,V_2^t,\m^t,\n^t$ unchanged.}\texshopboxnt{\pn 
\nitem{(d)} {\bf Policy improvement for the maximizer}: For all $x_2\in X_2$, set
$$J_2^{t+1}(x_2)=V_2^{t+1}(x_2)=\max_{v\in V(x_2)}H_2\big(x_2,v,\min[V_1^t,J_1^t]\big),\xdef\polimprmax{\lab}\eqnum\show{oneo}$$
set $\n^{t+1}(x_2)$ to a control $v\in V(x_2)$ that attains the above maximum,
and leave $J_1^t,V_1^t,\m^t$ unchanged.
}

\xdef\exampleminimaxcontrolpartialalg{\exampl}\examplnum\show{myexample}

\beginexample{\exampleminimaxcontrolpartialalg\ (Our PI Algorithm for Minimax Control - Explicit Separation of the Players)}\pn Consider the minimax control problem with explicit separation of the two players of Example \exampleminimaxcontroltwo, which involves the dynamic system 
 $x_{1,k}\in X_1$ and $x_{2,k}\in X_2$, and they evolve according to
  $$x_{2,k+1}=f_1(x_{1,k},u_k),\qquad x_{1,k+1}=f_2(x_{2,k+1},v_k),\qquad k=0,1,\ldots,$$
[cf.\ Eq.\ \minmaxone]. The Bellman equation for this problem can be broken down into the two equations \minmaxfive:
$$J_1^*(x_1)=\inf_{u\in U(x_1)}\Big[g_1(x_1,u)+\a J_2^*\big(f_1(x_1,u)\big)\Big],\qquad J_2^*(x_2)=\sup_{v\in V(x_2)}\Big[g_2(x_2,v)+\a J_1^*\big(f_2(x_2,v)\big)\Big].$$

In the context of this problem, the four operations \polevalmin-\polimprmax\ of our PI algorithm take the following form:
\nitem{(a)} {\bf Single value iteration for policy evaluation for the minimizer}: For all $x_1\in X_1$, set
$$J_1^{t+1}(x_1)=g_1\big(x_1,\m^t(x_1)\big)+\a \max\Big[V_2^t\big(f_1(x_1,\m^t(x_1))\big),J_2^t\big(f_1(x_1,\m^t(x_1))\big)\Big],\eqnum\show{oneo}$$
and leave $J_2^t,V_1^t,V_2^t,\m^t,\n^t$ unchanged.
\nitem{(b)} {\bf Policy improvement for the minimizer}: For all $x_1\in X_1$, set
$$J_1^{t+1}(x_1)=V_1^{t+1}(x_1)=\min_{u\in U(x_1)}\Big[g_1(x_1,u)+\a \max\big[V_2^t\big(f_1(x_1,u)\big),J_2^t\big(f_1(x_1,u)\big)\big]\Big],\eqnum\show{oneo}$$
set $\m^{t+1}(x_1)$ to a control $u\in U(x_1)$ that attains the above minimum,
and leave $J_2^t,V_2^t,\n^t$ unchanged.
\nitem{(c)} {\bf Single value iteration for policy evaluation of the maximizer}: For all $x_2\in X_2$ and $v\in V(x_2)$, set
$$J_2^{t+1}(x_2)=g_2\big(x_2,\n^t(x_2)\big)+\a \min\Big[V_1^t\big(f_2(x_2,\n^t(x_2))\big),J_1^t\big(f_2(x_2,\n^t(x_2))\big)\Big],\eqnum\show{oneo}$$
and leave $J_1^t,V_1^t,V_2^t,\m^t,\n^t$ unchanged.
\nitem{(d)} {\bf Policy improvement for the maximizer}: For all $x_2\in X_2$, set
$$J_2^{t+1}(x_2)=V_2^{t+1}(x_2)=\max_{v\in V(x_2)}\Big[g_2(x_2,v)+\a \min\big[V_1^t\big(f_2(x_2,v)\big),J_1^t\big(f_2(x_2,v)\big)\big]\Big],\eqnum\show{oneo}$$
set $\n^{t+1}(x_2)$ to a control $u\in V(x_2)$ that attains the above maximum,
and leave $J_1^t,V_1^t,\m^t$ unchanged.
\endexample

\vskip-1pc

\xdef\examplefourpi{\exampl}\examplnum\show{myexample}

\beginexample{\examplefourpi\ (Our PI Algorithm for Markov Games)}Let us consider the Markov game formulation of Example \examplefour.
Our PI algorithm with $x_1$, $x_2$, $H_1$, and $H_2$ defined earlier, can be implemented by storing $J_1^t,V_1^t$ as the real numbers $J_1^t(x)$ and $V_1^t(x)$, $x\in X,$
and by storing and representing the piecewise linear functions $J_2^t,V_2^t$ using the $m$ columns of the $n\times m$ matrices 
$$A(x)+\a\b\sum_{y\in X}Q_{xy}\min\big[V_1^t(y),J_1^t(y)\big],\qquad x\in X;\xdef\markovmatrix{\lab}\eqnum\show{oneo}$$
cf.\ Eq.\ \markovbeltwo. None of the operations \polevalmin-\polimprmax\ require the solution of a Markovian decision problem as in the Hoffman-Karp algorithm. This is  similar to the Pollatschek and Avi-Itzhak algorithm.  

More specifically, the policy evaluation \polevalmin\ for the minimizer takes the form
$$J_1^{t+1}(x)= {1\over \b}\max\Big[V_2^{t+1}\big(x,\m^t(x)\big),J_2^{t+1}\big(x,\m^t(x)\big)\Big],\qquad \hbox{for all }x\in X,\xdef\polevalminmarkov{\lab}\eqnum\show{oneo}$$
while the policy improvement \polimprmin\ for the minimizer takes the form
$$J_1^{t+1}(x)=V_1^{t+1}(x)={1\over \b}\min_{u\in U}\,\max\big[V_2^{t+1}(x,u),J_2^{t+1}(x,u)\big],\qquad \hbox{for all }x\in X.\eqnum\show{oneo}$$
The policy evaluation \polevalmax\ for the maximizer takes the form
$$J_2^{t+1}(x,u)= u'\left(A(x)+\a\b\sum_{y\in X}Q_{xy}\min\big[V_1^t(y),J_1^t(y)\big]\right)\big(\n^t(x)\big),\qquad \hbox{for all }x\in X,\,u\in U,\eqnum\show{oneo}$$
while the policy improvement \polimprmax\ for the maximizer takes the form
$$\eqalign{J_2^{t+1}(x,u)&=V_2^{t+1}(x,u)\cr
&=\max\Bigg\{ u'\left(A(x)+\a\b\sum_{y\in X}Q_{xy}\min\big[V_1^t(y),J_1^t(y)\big]\right)(1),\ldots,\cr
&\ \ \ \ \ \ \ \ \ \ \ \ \ \ \ \ \ \ \ \ u'\left(A(x)+\a\b\sum_{y\in X}Q_{xy}\min\big[V_1^t(y),J_1^t(y)\big]\right)(m)\Bigg\},\qquad \hbox{for all }x\in X,\,u\in U,\cr}\xdef\polimprmaxmarkov{\lab}\eqnum\show{oneo}$$
where 
$$\left(A(x)+\a\b\sum_{y\in X}Q_{xy}\min\big[V_1^t(y),J_1^t(y)\big]\right)(j)$$ 
is the $j$th column of the $n\times m$ matrix \markovmatrix. 

Again it can be seen that except for the extra memory storage to maintain $V_1^t$ and $V_2^t$, the preceding PI algorithm \polevalminmarkov-\polimprmaxmarkov\ requires roughly similar/comparable computations to the ones of the ``naive"  optimistic PI algorithm  \polevalPoAoptone-\polimprPoAtwofinal, when applied to the Markov game model.
\endexample
 \vskip-1pc

\subsubsection{Discussion of our Algorithm}

\pn Let us now provide a discussion of some of the properties of our PI algorithm \polevalmin-\polimprmax. We first note that except for the extra memory storage to maintain $V_1^t$ and $V_2^t$, the algorithm  requires roughly similar/comparable computations  to the ones of the ``naive"  optimistic PI algorithm  \polevalPoAoptone-\polimprPoAtwofinal. Note also that by performing a large number of value iterations of the form \polevalmin\ or \polevalmax\ we obtain an algorithm that involves nearly exact policy evaluation, similar to the ``naive"  nonoptimistic PI algorithm \polevalmarkovPoAone-\polimprPoAtwo.

Mathematically, under the contraction assumption to be introduced in the next section, our algorithm \polevalmin-\polimprmax\ avoids the oscillatory behavior illustrated in Fig.\ 2.1 because it embodies a policy-dependent sup-norm contraction, which has a {\it uniform fixed point\/}, the pair $(\jstar_1,\jstar_2)$, regardless of the policies. This is the essence of the key Prop.\ 4.2, which will be shown in the next section. 

Aside from this mathematical insight, one may gain intuition into the mechanism of our algorithm \polevalmin-\polimprmax, by comparing it with the optimistic version of the ``naive"  optimistic PI algorithm \polevalPoAoptone-\polimprPoAtwofinal. Our algorithm  \polevalmin-\polimprmax\ involves additionally the functions $V_1^t$ and $V_2^t$, which are changed only during the policy improvement operations, and tend to provide a guarantee against oscillatory behavior. In particular, since 
$$\max[V_2^t,J_2^t]\ge J_2^t,$$
the iterations of the minimizer in our algorithm, \polevalmin\ and \polimprmin, are more ``pessimistic" about the choices of the maximizer than the iterates of the minimizer in the ``naive" PI iterates \polevalPoAoptone\ and \polevalPoAopttwo. Similarly, since
$$\min[V_1^t,J_1^t]\le J_1^t,$$
 the iterations of the maximizer in our algorithm, \polevalmax\ and \polimprmax, are more ``pessimistic"  than the iterates of the maximizer in the naive PI iterates \polevalPoAoptone\ and \polevalPoAopttwo. As a result {\it the use of $V_1^t$ and $V_2^t$ in our PI algorithm makes it more conservative\/}, and mitigates the oscillatory swings that are illustrated in Fig.\ 2.1.
 
Let us also note that the use of the functions $V_1$ and $V_2$ in our algorithm  \polevalmin-\polimprmax\ may slow down the algorithmic progress relative to the (nonconvergent) ``naive" algorithm \polevalmarkovPoAone-\polimprPoAtwo. To remedy this situation an interpolation device has been suggested in the paper [BeY10] (Section V), which roughly speaking interpolates between the two algorithms, while still guaranteeing the algorithm's convergence; see also [Ber18], Section 2.6.3. Basically, such a device makes the algorithm less ``pessimistic," as it guards against nonconvergence, and it can similarly be used in our algorithm \polevalmin-\polimprmax.

In the next section, we will show convergence of our PI algorithm \polevalmin-\polimprmax\ with a line of proof that can be summarized as follows. Using a contraction argument, based on an assumption to be introduced shortly, we show that the sequences $\{V_1^t\}$ 
and $\{V_2^t\}$ converge to some functions $V^*_1\in B(X_1)$ and $V^*_2\in B(X_2)$, respectively. From the policy improvement operations \polimprmin\ and \polimprmax\ it will then follow that  the sequences $\{J_1^t\}$ and $\{J_2^t\}$ converge to the same functions $V^*_1$ and $V^*_2$, respectively, so that $\min[V_1^t,J_1^t]$ and $\max[V_2^t,J_2^t]$ converge to $V^*_1$ and $V^*_2$, respectively, as well. Using the continuity of $H_1$ and $H_2$ (a consequence of our contraction assumption), it follows from Eqs.\ \polimprmin\ and \polimprmax\ that  $(V^*_1,V^*_2)$ is the fixed point of $H_1$ and $H_2$ [in the sense of Eq.\ \doublefix], and hence is also equal to $(\jstar_1,\jstar_2)$ [cf.\ Eq.\ \doublefix]. Thus we finally obtain convergence:  
$V_1^t\to\jstar_1$, $J_1^t\to\jstar_1$, $V_2^t\to\jstar_2$, $J_2^t\to\jstar_2.$

\vskip-1.pc

\section{Convergence Analysis of our PI Algorithm}
\mark{Convergence Analysis of our PI Algorithm}
\vskip-1.5pc
\pn For each $\m\in{\cal M}$, we consider the operator $T_{1,\m}$ that maps a function $J_2\in B(X_2)$ into the function of $x_1$ given by
$$(T_{1,\m}J_2)(x_1)=H_1\big(x_1,\m(x_1),J_2\big),\qquad x_1\in X_1.\eqnum\show{oneo}$$
Also for each $\n\in{\cal N}$, we consider the operator $T_{2,\n}$ that maps a function $J_1\in B(X_1)$ into the function of $x_2$ given by
$$(T_{2,\n}J_1)(x_2)=H_2\big(x_2,\n(x_2),J_1\big),\qquad  x_2\in X_2.\eqnum\show{oneo}$$
We will also consider the operator $T_{\m,\n}$ that maps a function $(J_1,J_2)\in B(X_1)\times B(X_2)$ into the function of $(x_1,x_2)\in X_1\times X_2$, given by
$$\big(T_{\m,\n}(J_1,J_2)\big)(x_1,x_2)=\big((T_{1,\m}J_2)(x_1),\,(T_{1,\n}J_1)(x_2)\big).\xdef\opermunu{\lab}\eqnum\show{oneo}$$
[Recall here that the norms on $B(X_1)$, $B(X_2)$, and $B(X_1)\times B(X_2)$ are given by Eqs.\ \weightednorms\ and  \supnormtwo.]

We will show convergence of our algorithm assuming the following.

\texshopbox{\pn{\bf Assumption 4.1: (Contraction Assumption)} Consider the operator $T_{\m,\n}$ given by Eq.\ \opermunu.
\nitem{(a)}  For all  $(\m,\n)\in{\cal M}\times{\cal N}$, and $(J_1,J_2)\in B(X_1)\times B(X_2)$, the function $T_{\m,\n}(J_1,J_2)$ belongs to $B(X_1)\times B(X_2)$.
\nitem{(b)} There exists an $\a\in(0,1)$ such that  for all  $(\m,\n)\in{\cal M}\times{\cal N}$, $T_{\m,\n}$ is a contraction mapping  of modulus $\a$ within $B(X_1)\times B(X_2)$.
}

It can be verified that the preceding assumption holds in the case of Example \exampleminimaxcontroltwo\ (discounted minimax control with explicit separation of the two players), in the case of a discounted Markov game (cf.\ Example \examplefour), and in the cases of the discounted minimax control problems of Examples \exampleminimaxcontrol\ and \exampleminimaxcontrolpartial. For this we need the scaling parameter $\b$ introduced in Examples \examplefour-\exampleminimaxcontrolpartial.

By writing the contraction property of Assumption 4.1 as 
$$\max\big\{\|T_{1,\m}J_2-T_{1,\m}J'_2\|_1,\,\|T_{2,\n}J_1-T_{2,\n}J'_1\|_2\big\}\le \a \max\big\{\|J_1-J'_1\|_1,\,\|J_2-J'_2\|_2\big\},\xdef\contractionproperty{\lab}\eqnum\show{oneo}$$
for all $J_1,J'_1\in B(X_1)$ and $J_2,J'_2\in B(X_2)$ [cf.\ the norm definition \supnormtwo], we have
$$\|T_{1,\m}J_2-T_{1,\m}J'_2\|_1\le \a \|J_2-J'_2\|_2,\qquad \hbox{for all }J_2,J'_2\in B(X_2),\xdef\contractionmuone{\lab}\eqnum\show{oneo}$$
and
$$\|T_{2,\n}J_1-T_{2,\n}J'_1\|_2\le \a \|J_1-J'_1\|_1,\qquad \hbox{for all }J_1,J'_1\in B(X_1);\eqnum\show{oneo}$$
[set $J_1=J'_1$ or $J_2=J'_2$, respectively, in Eq.\ \contractionproperty]. From these relations, we obtain\footnote{\dag}{\ninepoint  
For a proof, we write Eq.\ \contractionmuone\ as
$$(T_{1,\m}J_2)(x_1)\le (T_{1,\m}J_2')(x_1)+\a\|J_2-J_2'\|_2\,\xi_1(x_1),\qquad (T_{1,\m}J_2')(x_1)\le (T_{1,\m}J_2)(x_1)+\a\|J_2-J_2'\|_2\,\xi_1(x_1),$$
for all $x_1\in X_1$. By taking infimum of both sides over $\m\in{\cal M}$, we obtain
$${\big|(T_1J_2)(x_1)-(T_1J'_2)(x_1)\big|\over \xi_1(x_1)}\le \a\|J_2-J_2'\|_2,$$
and by taking supremum over $x_1\in X_1$, the desired relation 
$\|T_{1}J_2-T_{1}J'_2\|_1\le \a \|J_2-J'_2\|_2$
 follows. The proof of the other relation, 
 $\|T_{2}J_1-T_{2}J'_1\|_2\le \a \|J_1-J'_1\|_1$, is similar.}
$$\|T_{1}J_2-T_{1}J'_2\|_1\le \a \|J_2-J'_2\|_2,\qquad \hbox{for all }J_2,J'_2\in B(X_2),\xdef\contractionone{\lab}\eqnum\show{oneo}$$
and
$$\|T_{2}J_1-T_{2}J'_1\|_2\le \a \|J_1-J'_1\|_1,\qquad \hbox{for all }J_1,J'_1\in B(X_1),\xdef\contractiontwo{\lab}\eqnum\show{oneo}$$
where
$$(T_1J_2)(x_1)=\inf_{\m\in{\cal M}}(T_{1,\m}J_2)(x_1)=\inf_{u\in U(x_1)}H_1(x_1,u,J_2),\qquad x_1\in X_1,$$
$$(T_2J_1)(x_2)=\sup_{\n\in{\cal N}}(T_{2,\n}J_1)(x_2)=\sup_{v\in V(x_2)}H_2(x_2,v,J_1),\qquad x_2\in X_2.$$
The relations \contractionone-\contractiontwo\ also imply that {\it the operator $T:B(X_1)\times B(X_2)\mapsto B(X_1)\times B(X_2)$ defined by
$$T(J_1,J_2)=(T_1J_2,T_2J_1),\xdef\minmaxt{\lab}\eqnum\show{oneo}$$
is a contraction mapping from $B(X_1)\times B(X_2)$ to $B(X_1)\times B(X_2)$ with modulus $\a$\/}. It follows that {\it $T$ has a unique fixed point $(\jstar_1,\jstar_2)\in B(X_1)\times B(X_2)$\/}. We will show that our algorithm yields in the limit this fixed point.

The following is our main convergence result [convergence here is meant in the sense of the norm \supnormtwo\ on $B(X_1)\times B(X_2)$]. Note that this result applies to any order and frequency of policy evaluations and policy improvements of the two players.

\xdef\propconvergence{\propn}\propnum\show{myproposition}

\texshopbox{\proposition{\propconvergence: (Convergence)} Let Assumption 4.1 hold, and assume that each of the four operations of the PI algorithm  \polevalmin-\polimprmax\ is performed infinitely often. Then the sequences $\big\{(J_1^t,J_2^t)\big\}$ and $\big\{(V_1^t,V_2^t)\big\}$ generated by the algorithm converge to $(\jstar_1,\jstar_2)$.
}

The proof is long but follows closely the steps of the proof for the single-player abstract DP case in Section 2.6.3 of the author's book [Ber18], which itself follows closely the steps of the corresponding proof for discounted Markovian decision problems given by Bertsekas and Yu [BeY12].

\subsubsection{An Extended Algorithm and its Convergence Proof}

\pn We first show the following lemma.

\texshopbox{\pn{\bf Lemma 4.1:} For all $(V_1,V_2), (J_1,J_2),(V'_1,V'_2), (J'_1,J'_2)\in B(X_1)\times B(X_2)$, we have
$$\big\|\min[V_1,J_1]-\min[V'_1,J'_1]\big\|_1\le \max\big\{ \|V_1-V'_1\|_1,\, \|J_1-J'_1\|_1\big\},\xdef\firstineq{\lab}\eqnum\show{oneo}$$
$$\big\|\max[V_2,J_2]-\min[V'_2,J'_2]\big\|_2\le \max\big\{ \|V_2-V'_2\|_2,\, \|J_2-J'_2\|_2\big\}.\xdef\secondineq{\lab}\eqnum\show{oneo}$$
}
\proof For every $x_1\in X_1$, we write
$${V_1(x_1)\over \xi_1(x_1)}\le {V'_1(x_1)\over \xi_1(x_1)}+\max\big\{ \|V_1-V'_1\|_1,\, \|J_1-J'_1\|_1\big\},$$
$${J_1(x_1)\over \xi_1(x_1)}\le {J'_1(x_1)\over \xi_1(x_1)}+\max\big\{ \|V_1-V'_1\|_1,\, \|J_1-J'_1\|_1\big\},$$
from which we obtain
$${\min\big\{V_1(x_1),J_1(x_1)\big\}\over \xi_1(x_1)}\le {\min\big\{V'_1(x_1),J'_1(x_1)\big\}\over \xi_1(x_1)}+\max\big\{ \|V_1-V'_1\|_1,\, \|J_1-J'_1\|_1\big\},$$
so that
$${\min\big\{V_1(x_1),J_1(x_1)\big\}-\min\big\{V'_1(x_1),J'_1(x_1)\big\}\over \xi_1(x_1)}\le \max\big\{ \|V_1-V'_1\|_1,\, \|J_1-J'_1\|_1\big\}.$$
By exchanging the roles of $(V_1,J_1)$ and $(V'_1,J'_1)$, and combining the two inequalities, we have
$${\Big|\min\big\{V_1(x_1),J_1(x_1)\big\}-\min\big\{V'_1(x_1),J'_1(x_1)\big\}\Big|\over \xi_1(x_1)}\le \max\big\{ \|V_1-V'_1\|_1,\, \|J_1-J'_1\|_1\big\},$$
and by taking the supremum over $x_1\in X_1$, we obtain Eq.\ \firstineq. We similarly prove Eq.\ \secondineq. \qed

We consider the spaces of bounded functions $Q_1(x_1,u)$ of $(x_1,u)\in X_1\times U$ and $Q_1(x_2,v)$ of $(x_2,v)\in X_2\times V$, with norms
$$\|Q_1\|_1=\sup_{x_1\in X_1,\,u\in U}{\big|Q_1(x_1,u)\big|\over \xi_1(x_1)},\qquad\|Q_2\|_2=\sup_{x_2\in X_2,\,v\in V}{\big|Q_2(x_2,v)\big|\over \xi_2(x_2)},\xdef\normtwo{\lab}\eqnum\show{oneo}$$
respectively, where $\xi_1$ and $\xi_2$ are the weighting functions that define the norm of $B(X_1)$ and $B(X_2)$ [cf.\ Eq.\ \weightednorms].  We denote these spaces by $B(X_1\times U)$ and $B(X_2\times V)$, respectively. Functions in these spaces have the meaning of {\it Q-factors for the minimizer and the maximizer\/}.

We next introduce  a new operator, denoted by $G_{\m,\n}$, which is parametrized by the policy pair $(\m,\n)$, and will be shown to have a common fixed point for all $(\m,\n)\in{\cal M}\times{\cal N}$, from which $(\jstar_1,\jstar_2)$ can be readily obtained. The operator  $G_{\m,\n}$ involves operations on Q-factor pairs  $(Q_1,Q_2)$ for  the minimizer and the maximizer, in addition to functions of state $(V_1,V_2)$, and is used define an ``extended" PI algorithm that operates over a larger function space than the one of Section 3. Once the convergence of this ``extended" PI algorithm is shown, the convergence of our algorithm of Section 3 will readily follow.

To define the operator  $G_{\m,\n}$, we note that it consists of four components, maps $B(X_1)\times B(X_2)\times B(X_1\times U)\times B(X_2\times V)$ into itself. It is given by
$$G_{\m,\n}(V_1,V_2,Q_1,Q_2)=\big(M_{1,\n}(V_2,Q_2),M_{2,\m}(V_1,Q_1),F_{1,\n}(V_2,Q_2),F_{2,\m}(V_1,Q_1)\big),\xdef\composite{\lab}\eqnum\show{oneo}$$
where the  functions
$M_{1,\n}(V_2,Q_2)$, $M_{2,\m}(V_1,Q_1)$, $F_{1,\n}(V_2,Q_2),$ and $F_{2,\m}(V_1,Q_1),$ 
are defined as follows:

\nitem{$\bullet$} $M_{1,\n}(V_2,Q_2)$: This is the function of $x_1$ given  by
$$\big(M_{1,\n}(V_2,Q_2)\big)(x_1)=\big(T_1\max[V_2,\hat Q_{2,\n}]\big)(x_1)=\inf_{u\in U(x_1)}H_1\big(x_1,u,\max[V_2,\hat Q_{2,\n}]\big),\xdef\minmapextone{\lab}\eqnum\show{oneo}$$
where $\hat Q_{2,\n}$ is the function of $x_2$ given by
$$\hat Q_{2,\n}(x_2)=Q_2\big(x_2,\n(x_2)\big).\eqnum\show{oneo}$$

\nitem{$\bullet$} $M_{2,\m}(V_1,Q_1)$: This is the function of $x_2$ given  by
$$\big(M_{2,\m}(V_1,Q_1)\big)(x_2)=\big(T_2\min[V_1,\hat Q_{1,\m}]\big)(x_2)=\sup_{v\in V(x_2)}H_2\big(x_2,v,\min[V_1,\hat Q_{1,\m}]\big),\xdef\minmapexttwo{\lab}\eqnum\show{oneo}$$
where $\hat Q_{1,\m}$ is the function of $x_1$ given by
$$\hat Q_{1,\m}(x_1)=Q_1\big(x_1,\m(x_1)\big).\eqnum\show{oneo}$$

\nitem{$\bullet$} $F_{1,\n}(V_2,Q_2)$: This is the function of $(x_1,u)$, given  by
$$F_{1,\n}(V_2,Q_2)(x_1,u)=H_1\big(x_1,u,\max[V_2,\hat Q_{2,\n}]\big).\eqnum\show{oneo}$$
 
\nitem{$\bullet$} $F_{2,\m}(V_1,Q_1)$: This is the function of $(x_2,v)$, given  by
$$F_{2,\m}(V_1,Q_1)(x_2,v)=H_2\big(x_2,v,\min[V_1,\hat Q_{1,\m}]\big).\xdef\fmumaptwo{\lab}\eqnum\show{oneo}$$
\smskip

Note that the four components of $G_{\m,\n}$ correspond to the four operations of our algorithm \polevalmin-\polimprmax. In particular,
\nitem{$\bullet$} $M_{1,\n}(V_2,Q_2)$ corresponds to policy improvement of the minimizer.
\nitem{$\bullet$} $M_{2,\m}(V_1,Q_1)$ corresponds to policy improvement of the maximizer.
\nitem{$\bullet$} $F_{1,\n}(V_2,Q_2)$ corresponds to policy evaluation of the minimizer.
\nitem{$\bullet$} $F_{2,\m}(V_1,Q_1)$ corresponds to policy evaluation of the maximizer.
\smskip

The key step in our convergence proof is to show that $G_{\m,\n}$ has a contraction property with respect to the norm on $B(X_1)\times B(X_2)\times B(X_1\times U)\times B(X_2\times V)$ given by
$$\big\|(V_1,V_2,Q_1,Q_2)\big\|=\max\big\{\| V_1\|_1,\| V_2\|_2,\| Q_1\|_1,\| Q_2\|_2\big\},\xdef\normone{\lab}\eqnum\show{oneo}$$
where $\| V_1\|_1$, $\| V_2\|_2$ are the weighted sup-norms of $V_1$, $V_2$, respectively, defined by Eq.\ \weightednorms, and $\| Q_1\|_1$, $\| Q_2\|_2$ are the weighted sup-norms of $Q_1$, $Q_2$, defined by Eq.\ \normtwo. Moreover, the contraction property is uniform, in the sense that {\it the fixed point of $G_{\m,\n}$ does not depend on $(\m,\n)$\/}. This means that {\it we can carry out iterations with 
$G_{\m,\n}$, while changing $\m$ and $\n$ arbitrarily between iterations, and still aim at the same fixed point\/}.
We have the following proposition.

\xdef\propunicontraction{\propn}\propnum\show{myproposition}

\texshopbox{\proposition{\propunicontraction: (Uniform Contraction)} Let Assumption 4.1 hold. Then for all $(\m,\n)\in{\cal M}\times{\cal N}$, the operator $G_{\m,\n}$ is a contraction mapping with modulus $\a$ with respect to the norm of Eqs.\ \normone, \weightednorms, and \normtwo. Moreover, the corresponding fixed point of $G_{\m,\n}$ is $(\jstar_1,\jstar_2,\qstar_1,\qstar_2)$ [independently of the choice of $(\m,\n)$], where $(\jstar_1,\jstar_2)$ is the fixed point of the mapping $T$ of Eq.\ \minmaxt, and $\qstar_1,\qstar_2$ are the functions defined by
$$\qstar_1 (x_1,u)=H_1(x_1,u,\jstar_2),\qquad x_1\in X_1,\,u\in U(x_1),\xdef\qstardefone{\lab}\eqnum\show{oneo}$$
$$\qstar_2 (x_2,v)=H_2(x_2,v,\jstar_1),\qquad x_2\in X_2,\,v\in V(x_2).\xdef\qstardeftwo{\lab}\eqnum\show{oneo}$$
}
 
 \proof We prove the contraction property of $G_{\m,\n}$ by breaking it down to four inequalities, which hold for all $(V_1,V_2), (V'_1,V'_2)\in B(X_1)\times B(X_2)$ and $(Q_1,Q_2),(Q'_1,Q'_2)\in B(X_1,U)\times B(X_2,V)$. In particular, we have
$$\eqalign{\big\|M_{1,\n}(V_2,Q_2)-M_{1,\n}(V'_2,Q'_2)\big\|_1&=\Big\|T_{1}\big(\max[V_2,\hat Q_{2,\n}]\big)-T_{1}\big(\max[V'_2,\hat Q'_{2,\n}]\big)\Big\|_1\cr
&\le \a\big\|\max[V_2,\hat Q_{2,\n}]_2-\max[V'_2,\hat Q'_{2,\n}]\big\|_2\cr
&\le \a\max\big\{\|V_2-V'_2\|_2,\, \|\hat Q_{2,\n}-\hat Q'_{2,\n}\|_2\big\}\cr
&\le \a\max\big\{\|V_2-V'_2\|_2,\, \|Q_2-Q'_2\|_2\big\}\cr
&\le \a\max\big\{ \|V_1-V'_1\|_1,\, \|Q_1-Q'_1\|_1,\, \|V_2-V'_2\|_2,\, \|Q_2-Q'_2\|_2\big\}\cr
&=\a\,\big\|(V_1,V_2,Q_1,Q_2)-(V'_1,V'_2,Q'_1,Q'_2)\big\|,\cr}\xdef\ineqone{\lab}\eqnum\show{oneo}$$
where the first equality uses the definitions of $M_{1,\n}(V_2,Q_2)$, $M_{1,\n}(V'_2,Q'_2)$ [cf.\  Eqs.\ \minmapextone\ and \minmapexttwo], the first inequality follows from Eq.\ \contractionmuone, the second inequality follows using Lemma 4.1, the third inequality follows from the definition of $\hat Q_{2,\n}$ and $\hat Q'_{2,\n}$, the last inequality is trivial, and the last equality follows from the norm definition \normone.
Similarly, we prove that
$$\big\|M_{2,\m}(V_1,Q_1)-M_{2,\m}(V'_1,Q'_1)\big\|_2\le \a\,\big\|(V_1,V_2,Q_1,Q_2)-(V'_1,V'_2,Q'_1,Q'_2)\big\|,\eqnum\show{oneo}$$
$$\big\|F_{1,\n}(V_2,Q_2)-F_{1,\n}(V'_2,Q'_2)\big\|_1\le \a\,\big\|(V_1,V_2,Q_1,Q_2)-(V'_1,V'_2,Q'_1,Q'_2)\big\|,\eqnum\show{oneo}$$
 $$\big\|F_{2,\m}(V_1,Q_1)-F_{2,\m}(V'_1,Q'_1)\big\|_2
 \le \a\,\big\|(V_1,V_2,Q_1,Q_2)-(V'_1,V'_2,Q'_1,Q'_2)\big\|.\xdef\ineqfour{\lab}\eqnum\show{oneo}$$

From the preceding relations \ineqone-\ineqfour, it follows that each of the four components of the maximization that comprises the norm 
$$\big\|G_{\m,\n}(V_1,V_2,Q_1,Q_2)-G_{\m,\n}(V'_1,V'_2,Q'_1,Q'_2)\big\|$$
[cf.\ Eq.\ \composite] is less or equal to
$$\a\,\big\|(V_1,V_2,Q_1,Q_2)-(V'_1,V'_2,Q'_1,Q'_2)\big\|.$$
Thus we have
$$\big\|G_{\m,\n}(V_1,V_2,Q_1,Q_2)-G_{\m,\n}(V'_1,V'_2,Q'_1,Q'_2)\big\|\le \a\,\big\|(V_1,V_2,Q_1,Q_2)-(V'_1,V'_2,Q'_1,Q'_2)\big\|,$$
which shows the desired contraction property of $G_{\m,\n}$.

In view of the contraction property just shown, the mapping $G_{\m,\n}$ has a unique fixed point for each $(\m,\n)\in{\cal M}\times{\cal N}$, which we denote by $(V_1,V_2,Q_1,Q_2)$ [with some notational abuse, we do not show the possible dependence of the fixed point on $(\m,\n)$]. In view of 
 Eqs.\ \composite-\fmumaptwo, this fixed point satisfies for all $x_1\in X_1$, $x_2\in X_2$, $(x_1,u)\in X_1\times U$, $(x_2,v)\in X_2\times V$, 
 $$V_1(x_1)=\inf_{u'\in U(x_1)}H_1\big(x_1,u',\max[V_2,\hat Q_{2,\n}]\big),\qquad V_2(x_2)=\sup_{v'\in V(x_2)}H_2\big(x_2,v',\min[V_1,\hat Q_{1,\m}]\big),\xdef\defqoneqone{\lab}\eqnum\show{oneo}$$
 $$Q_1(x_1,u)=H_1\big(x_1,u,\max[V_2,\hat Q_{2,\n}]\big),\qquad Q_2(x_2,v)=H_2\big(x_2,v,\min[V_1,\hat Q_{1,\m}]\big).\xdef\defqoneqtwo{\lab}\eqnum\show{oneo}$$
 By comparing the preceding two relations, it follows that for all $x_1\in X_1$, $x_2\in X_2$, 
 $$V_1(x_1)\le Q_1(x_1,u),\quad \hbox{for all }x_1,\,u\in U(x_1),$$
$$V_2(x_2)\ge Q_2(x_2,v),\quad \hbox{for all }x_2,\,v\in V(x_2),$$
 which implies that 
 $$\min[V_1,\hat Q_{1,\m}]=V_1,\qquad \max[V_2,\hat Q_{2,\n}]=V_2.$$
 Using Eq.\ \defqoneqone, this in turn shows that 
 $$V_1(x_1)=\inf_{u\in U(x_1)}H_1(x_1,u,V_2),\qquad V_2(x_2)=\sup_{v\in V(x_2)}H_2(x_2,v,V_1).$$
 Thus, independently of $(\m,\n)$, $(V_1,V_2)$ is the unique fixed point of the contraction mapping $T$ of Eq.\ \minmaxt, which is $(\jstar_1,\jstar_2)$. Moreover from Eq.\ \defqoneqtwo, we have that $(Q_1,Q_2)$ is precisely $(Q_1^*,Q_2^*)$ as given by Eqs.\ \qstardefone\ and \qstardeftwo. This shows that, independently of $(\m,\n)$, the fixed point of $G_{\m,\n}$ is $(\jstar_1,\jstar_2,Q_1^*,Q_2^*)$, and proves the desired result. \qed
 
 The preceding proposition implies the convergence of the ``extended" algorithm, which at each iteration $t$ applies one of the four components of $G_{\m^t,\n^t}$ evaluated at the current iterate $(V_1^t,V_2^t,Q_1^t,Q_2^t,\m^t,\n^t)$, and updates this iterate accordingly. 
This algorithm is well-suited for the calculation of both $(\jstar_1,\jstar_2)$ and $(\qstar_1,\qstar_2)$. However, since we are just interested to calculate $(\jstar_1,\jstar_2)$, a simpler and more efficient algorithm is possible, which is in fact our PI algorithm based on the four operations \polevalmin-\polimprmax. To this end, we
observe that the algorithm that updates $(V_1^t,V_2^t,Q_1^t,Q_2^t,\m^t,\n^t)$ can be operated so that it does not require the maintenance of the full Q-factor functions $(Q_1^t,Q_2^t)$. The reason is that the values $Q_1^t(x_1,u)$ and $Q_2^t(x_2,v)$ with $u\ne \m^t(x_1)$ and $v\ne \n^t(x_2)$, do not appear in the calculations, and hence we need only the values $\hat Q^t_{1,\m^t}(x_1)$ and $\hat Q^t_{2,\n^t}(x_2)$, which we store in functions $J^t_1$ and $J^t_2$, i.e., we set
$$J^t_1(x_1)=\hat Q^t_{1,\m^t}(x_1)=Q^t_1\bl(x_1,\m^t(x)\br),$$
$$J^t_2(x_2)=\hat Q^t_{2,\n^t}(x_2)=Q^t_2\bl(x_2,\n^t(x_2)\br).$$
Once we do that, the resulting algorithm is precisely our PI algorithm \polevalmin-\polimprmax.

In summary,  our PI algorithm \polevalmin-\polimprmax\  that updates $(V_1^t,V_2^t,J_1^t,J_2^t,\m^t,\n^t)$ is a reduced space implementation of the asynchronous fixed point algorithm that updates $(V_1^t,V_2^t,Q_1^t,Q_2^t,\m^t,\n^t)$ using the uniform contraction mapping $G_{\m^t,\n^t}$, with the identifications 
$$J_1^t=\hat Q_{1,\m^t},\qquad J_2^t=\hat Q_{2,\n^t}.$$
This proves its convergence as stated in Prop.\ \propconvergence. 

\vskip-1.5pc

\section{Reinforcement Learning Algorithms}
\mark{Reinforcement Learning Algorithms}
\vskip-2pc

\pn  Our algorithm of Section 3 involves exact implementation without function approximations, and thus is not suitable for large state and control spaces. An important research direction is approximate implementations based on our PI algorithmic structure of Section 3, whereby we use approximation in value space with cost function approximations obtained through reinforcement learning methods. An interesting algorithmic approach is aggregation with representative states, as described in the book [Ber19b] (Section 6.1).

 In particular, let us consider the minimax formulation of Example \exampleminimaxcontroltwo\ and Eqs.\ \minmaxone-\minmaxfive, which involves separate state spaces $X_1$ and $X_2$ for the minimizer and the maximizer, respectively. 
 In the aggregation  with representative states formalism, we execute our PI algorithm over reduced versions of the spaces $X_1$ and $X_2$. In particular, we  discretize $X_1$ and $X_2$ by using  suitable finite collections  of representative states $\tl X_1\subset X_1$ and $\tl X_2\subset X_2$, and construct a lower-dimensional aggregate problem. The typical stage involves transitions between representative states, with intermediate artificial transitions $x_1\to \tl x_1$ and  $x_2\to \tl x_2$, which involve randomization with aggregation probabilities $\phi_{x_1\skew3\tl x_1}$ and $\phi_{x_2\skew3\tl x_2}$, respectively; see Fig.\ 5.1.

\midinsert
\centerline{\hskip0pc\epsfxsize = 5in \epsfbox{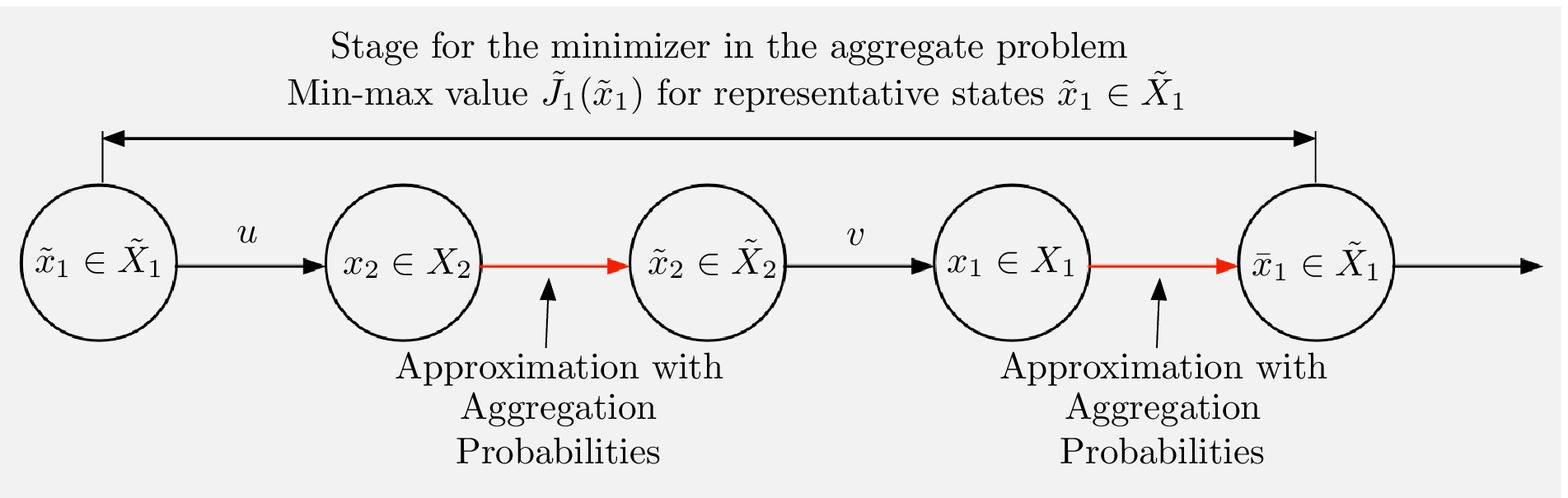}}
\vskip-1pc
\fig{0pc}{5.1.} {Schematic illustration of an aggregation framework that is patterned after the sequence of events of the multistage process of Fig.\ 3.1. The aggregate problem is specified by a finite subset of representative states $\skew3\tl X_1\subset X_1$, a finite subset of representative states $\skew3\tl X_2\subset X_2$, and aggregation probabilities for passing from states $x_2\in X_2$ to representative states $\skew3\tl x_2\in \skew3\tl X_2$, and for passing from states $x_1\in X_1$ to representative states $\skew3\tl x_1\in \skew3\tl X_1$. A stage starts at  a representative state $\skew3\tl x_1\in \skew3\tl X_1$ and ends at some other representative state $\bar x_1\in \skew3\tl X_1$, by going successively through a state $x_2\in X_2$ under the influence of the minimizer's choice $u\in U(\skew3\tl x_1)$, then to a representative state $\skew3\tl x_2\in \skew3\tl X_2$ using aggregation probabilities $\phi_{x_2\skew3\tl x_2}$ (i.e., the transition $x_2\to\tl x_2$ takes place with probability $\phi_{x_2\skew3\tl x_2}$), then to a state $x_1\in X_1$ under the influence of the maximizer's choice $v\in V(\skew3\tl x_2)$, and finally to $\bar x_1\in \skew3\tl X_1$  using aggregation probabilities $\phi_{x_1\bar x_1}$  (the transition $x_1\to\bar x_1$ takes place with probability $\phi_{x_1\bar x_1}$). The transitions $\tl x_1\to x_2$ and $\tl x_2\to x_1$ produce costs $g_1(\tl x_1,u)$ and $g_2(\tl x_2,v)$, respectively [cf.\ Eq.\ \minmaxthree]. The aggregation probabilities $\phi_{x_2\skew3\tl x_2}$ and $\phi_{x_1\bar x_1}$ can be arbitrary. However, their choice affects the min-max and max-min functions of the aggregate problem. 

We can solve the aggregate problem by using simulation-based versions of our PI algorithm \polevalmin-\polimprmax\ of Section 3 to obtain the min-max and max-min functions $\skew5\tl J_1(\skew3\tl x_1)$ and $\skew5\tl J_2(\skew3\tl x_2)$ at all the representative states $\skew3\tl x_1\in \skew3\tl X_1$ and  $\skew3\tl x_2\in \skew3\tl X_2$, respectively [cf.\ [Ber19b] (Chapter 6)]. Then, min-max and max-min function approximations are computed from
$$\skew5\tl J_1(x_1)=\sum_{\skew5\tl x_1\in \skew3\tl X_1}\phi_{x_1\skew3\tl x_1}\skew5\tl J_1(\skew3\tl x_1),\qquad \skew5\tl J_2(x_2)=\sum_{\skew3\tl x_2\in \skew3\tl X_2}\phi_{x_2\skew3\tl x_2}\skew5\tl J_2(\skew3\tl x_2).$$
Suboptimal decision choices by the minimizer and the maximizer are then obtained from the one-step lookahead optimizations 
$$\min_{u\in U(x_1)}H_1(x_1,u,\skew5\tl J_2),\qquad \max_{v\in V(x_2)}H_2(x_2,v,\skew5\tl J_1).$$
See the book [Ber19b] (Section 6.1) and the paper [Ber18b] for a detailed accounting of the aggregation approach with representative states for single-player infinite horizon DP.
}
\endinsert

The structure of the aggregate problem is amenable to a DP formulation, and as a result, it can be solved by using simulation-based versions of the PI methods of Section 3 [we refer to the book [Ber19b] (Chapter 6) for more details]. The cost function approximations thus obtained, call them $\tl J_1$, $\tl J_2$, are used in the one-step lookahead minimization 
 $$\min_{u\in U(x_1)}H_1(x_1,u,\tl J_2),$$
 to obtain a suboptimal  minimizer's policy, and  in the one-step lookahead maximization 
 $$\max_{v\in V(x_2)}H_2(x_2,v,\tl J_1),$$
 to obtain a suboptimal  maximizer's policy.

The aggregation with representative states approach has the advantage that it maintains the DP structure of the original minimax problem. This allows the use of our PI methods of Section 3, with convergence guaranteed by the results of Section 4. Another aggregation approach that can be similarly used within our context, is hard aggregation, whereby the state spaces $X_1$ and $X_2$ are partitioned into subsets that form aggregate states; see [Ber18b], [Ber18c], [Ber19b]. Other reinforcement learning methods, based for example on the use of neural networks, can also be used for approximate implementation of our PI algorithms. However, their convergence properties are problematic, in the absence of additional assumptions. The papers by Bertsekas and Yu ([BeY12], Sections 6 and 7), and by Yu and Bertsekas [YuB13] (Section 4), also describe alternative simulation-based approximation possibilities that may serve as a starting point for minimax PI algorithms with function approximation.

\vskip-1.5pc

\section{Conclusions and Extensions}
\mark{Conclusions and Extensions}

\vskip-2.pc

\pn In this paper, we have discussed PI algorithms that are specifically tailored to sequential zero-sum games and minimax problems with a contractive abstract DP structure. Our algorithms of Section 3 resolve the long-standing convergence difficulties of the Pollatschek and Avi-Itzhak PI algorithm [PoA69], and allow an asynchronous implementation, whereby the policy evaluation and policy improvement operations can be done in any order and with different frequencies. Moreover, our algorithms find simultaneously the min-max and the max-min values, and they are suitable for Markov zero-sum game problems, as well as for minimax control problems involving set-membership uncertainty.

While we have not addressed in detail the issue of asynchronous distributed implementation in a multiprocessor system, our algorithm admits such an implementation, as has been discussed for its  single-player counterparts in the papers by Bertsekas and Yu [BeY10], [BeY12], [YuB13], and also in a more abstract form in the author's books [Ber12], [Ber18], [Ber20]. In particular, there is a
 highly parallelizable and convergent distributed implementation, which is based on state space partitioning, and asynchronous policy evaluation and policy improvement operations within each set of the partition. The key idea, which forms the core of asynchronous DP algorithms [Ber82], [Ber83]  (see also the books [BeT89], [Ber12], [Ber18], [Ber20]) is that the mapping $G_{\m,\n}$ of Eq.\ \composite\ has two components  {\it for every state\/} (policy evaluation and policy improvement) for the minimizer and two corresponding components for every state for the maximizer. Because of the uniform sup-norm contraction property of $G_{\m,\n}$, iterating with any one of these components, and at any single state, does not impede the progress made by iterations with the other components, while making eventual progress towards the solution. 
 
In view of its asynchronous convergence capability, our framework is also suitable for on-line implementations where policy improvement and evaluations are done at only one state at a time. In such implementations, the algorithm performs a policy improvement at a single state, followed by a number of policy evaluations at other states, with the current policy pair $(\m^t,\n^t)$ evaluated at only one state $x$ at a time, and the cycle is repeated. One may select states cyclically for policy improvement, but there are alternative possibilities, including the case where states are selected on-line as the system operates. An on-line PI algorithm of this type, which may also be operated as a rollout algorithm (a control selected by a policy improvement at each encountered state), was given recently in the author's paper [Ber21], and can be straightforwardly adapted to the minimax and Markov game cases of this paper.

Other algorithmic possibilities, also discussed in the works just noted, involve the presence of ``communication delays" between processors, which roughly means that the iterates generated at some processors may involve iterates of other processors that are out-of-date. This is possible because the asynchronous convergence line of analysis framework of [Ber83] in combination with the uniform weighted sup-norm contraction property of Prop.\ \propunicontraction\ can tolerate the presence of such delays.  Implementations that involve forms of stochastic sampling have also been given in the papers [BeY12], [YuB13]. We defer further discussion along this line for a future report.

An important issue for efficient implementation of our algorithm is the relative frequency of policy improvement and policy evaluation operations. If a very large number of contiguous policy evaluation operations, using the same policy pair $(\m^t,\n^t)$, is done between policy improvement operations, the policy evaluation is nearly exact. Then the algorithm's behavior is essentially the same as the one of  the nonoptimistic algorithm where policy evaluation is done according to
$$J_{1,\m^t,\n^t}(x_1)=H_1\Big(x_1,\m^t(x_1),\max\big[V_2^t,J_{2,\m^t,\n^t}\big]\Big),\qquad x_1\in X_1,$$
$$J_{2,\m^t,\n^t}(x_2)=H_2\Big(x_2,\n^t(x_2),\min\big[V_1^t,J_{1,\m^t,\n^t}\big]\Big),\qquad x_2\in X_2,$$
cf.\ Eqs.\ \polevalmarkovPoAone-\polevalmarkovPoAtwo\ (in the context of Markovian decision problems, this type of policy evaluation involves the solution of an optimal stopping problem; cf.\ the paper [BeY12]). Otherwise the policy evaluation is inexact/optimistic, and in the extreme case where only one policy evaluation is done between policy improvements, the algorithm resembles a value iteration method. Based on experience with optimistic PI, it appears that the optimal number of policy evaluations between policy improvements should be substantially larger than one, and should also be problem-dependent.

We mention the possibility of extensions to other related minimax and Markov game problems. In particular, the treatment of undiscounted problems that involve a termination state can be patterned after the distributed asynchronous PI algorithm for stochastic shortest path problems by Yu and Bertsekas [YuB13], and will be the subject of a separate report.  A related area of investigation is on-line algorithms applied to robust shortest path planning problems, where the aim is to reach a termination state at minimum cost and against the actions of an antagonistic opponent. The author's paper [Ber19c] (see also the book [Ber18], Section 3.5.3) has provided analysis and algorithms, some of the PI type, for these minimax versions of shortest path problems, and has given many references of related works. Still our PI algorithm of Section 3, appropriately extended, offers some substantial advantages within the shortest path context, in both a serial and a distributed computing environment. 

Note that a sequential minimax problem with a finite horizon may be viewed as a simple special case of an infinite horizon problem with a termination state. The PI algorithms of the present paper are directly applicable and can be simply modified for such a problem. In conjunction with function approximation methods, such as the aggregation method described earlier, they may provide an attractive alternative to exact, but hopelessly time-consuming solution approaches. 

For an interesting class of finite horizon problems, consider a two-stage ``robust" version of stochastic programming, patterned after Example \exampleminimaxcontrol\ and Eq.\ \syseq. Here, at an initial state $x_0$, the decision maker/minimizer  applies a decision $u_0\in U(x_0)$, an antagonistic nature chooses $v_0\in V(x_0,u_0)$, and a random disturbance $w_0$ is generated according to a probability distribution than depends on $(x_0,u_0,v_0)$. A cost $g_0(x_0,u_0,v_0,w_0)$ is then incurred and the next state 
$$x_1=f(x_0,u_0,v_0,w_0)$$
is generated. Then the process is repeated at the second stage, with $(x_1,u_1,v_1,w_1)$ replacing $(x_0,u_0,v_0,w_0)$, and finally a terminal cost $G_2(x_2)$ is incurred where
$$x_2=f(x_1,u_1,v_1,w_1).$$
Here the decision maker aims to minimize the expected total cost assuming a worst-case selection of $(v_0,v_1)$. The maximizing choices $(v_0,v_1)$ may have a variety of problem-dependent interpretations, including prices affecting the costs $g_0$, $g_1$, $G_2$, and forecasts affecting the probability distributions of the disturbances $(w_0,w_1)$. The distributed asynchronous PI algorithm of Section 3 is easily modified for this problem, and similarly can be interpreted as  Newton's method for solving a two-stage version of Bellman's equation. Exact solution of the problem may be a daunting computational task, but a satisfactory suboptimal solution, along the lines of Section 5, using approximation in value space with function approximation based on aggregation may prove feasible.

Finally, let us note a theoretical use of our line of analysis that is based on  uniform contraction properties. It may form the basis for a rigorous mathematical treatment of PI algorithms in stochastic two-player DP models that  involve universally measurable policies. We refer to  the paper by Yu and Bertsekas [YuB15], where the associated issues of validity and convergence of PI methods for single-player problems have been addressed using algorithmic ideas that are closely related to the ones of the present paper.

\vskip-1. pc

\def\ref{\vskip1pt\pn}

\section{References}
\mark{References}

\vskip-2pc

\ref[AkG01] Akian, M., and Gaubert, S., ``Policy Iteration for Perfect Information Stochastic Mean Payoff Games with
Bounded First Return Times is Strongly Polynomial," arXiv preprint arXiv:1310.4953.

\ref[BFH86] Breton, M., Filar, J.\ A., Haurie, A., and Schultz, T.\ A., 1986.\ ``On the Computation of Equilibria in Discounted Stochastic Dynamic Games," in Dynamic Games and Applications in Economics, Springer,  pp.\ 64-87.

\ref [BeS78]  Bertsekas, D.\ P., and Shreve, S.\ E., 1978.\  Stochastic Optimal
Control:  The Discrete Time Case, Academic Press, N.\ Y.;
republished by Athena Scientific, Belmont, MA, 1996 (can be  downloaded from
the author's website).

\ref [BeT89]  Bertsekas, D.\ P., and Tsitsiklis, J.\ N., 1989.\ Parallel and
Distributed Computation: Numerical Methods, Prentice-Hall, Engl.\ 
Cliffs, N.\ J.\  (can be  downloaded from
the author's website).

\ref [BeT91]  Bertsekas, D.\ P., and Tsitsiklis, J.\ N., 1991.\ ``An Analysis of
Stochastic Shortest Path Problems,"
Math.\ Operations Research, Vol.\ 16, pp.\ 580-595.

\ref[BeY10] Bertsekas, D.\ P., and Yu, H., 2010.\ ``Asynchronous Distributed Policy Iteration in Dynamic Programming,"  Proc.\ of Allerton Conf.\ on Communication, Control and Computing,  Allerton Park, Ill, pp.\ 1368-1374.

\ref[BeY12] Bertsekas, D.\ P., and Yu, H., 2012.\ ``Q-Learning and Enhanced Policy Iteration in Discounted 
Dynamic Programming,"  Math.\ of Operations Research, Vol.\ 37, pp.\ 66-94.

\ref [Ber82] Bertsekas, D.\ P., 1982.\  ``Distributed Dynamic Programming," IEEE
Trans.\ Aut.\  Control, Vol.\ AC-27, pp.\ 610-616.

\ref[Ber83] Bertsekas, D.\ P., 1983.\ ``Asynchronous Distributed Computation of Fixed
Points," Math. Programming, Vol.\ 27, pp.\ 107-120.

\ref[Ber12] Bertsekas, D.\ P., 2012.\
Dynamic Programming and Optimal Control, Vol.\ II,  4th Ed., Athena Scientific, Belmont, MA.

\ref[Ber16] Bertsekas, D.\ P., 2016.\
Nonlinear Programming, Athena Scientific, Belmont, MA.

\ref[Ber18a] Bertsekas, D.\ P., 2018.\ Abstract Dynamic Programming, 2nd Ed., Athena Scientific, Belmont, MA  (can be downloaded from the author's website).

\ref[Ber18b] Bertsekas, D.\ P., 2018.\ ``Feature-Based Aggregation and Deep Reinforcement Learning: A Survey and Some New Implementations," Lab.\ for Information and Decision Systems Report, MIT; arXiv preprint arXiv:1804.04577; IEEE/CAA Journal of Automatica Sinica, Vol.\ 6, 2019, pp.\ 1-31.

\ref[Ber18c] Bertsekas, D.\ P., 2018.\ ``Biased Aggregation, Rollout, and Enhanced Policy Improvement for Reinforcement Learning," Lab.\ for Information and Decision Systems Report, MIT; arXiv preprint arXiv:1910.02426.

\ref[Ber19a] Bertsekas, D.\ P., 2019.\ ``Affine Monotonic and Risk-Sensitive Models in Dynamic Programming," IEEE Transactions on Aut.\ Control, Vol.\ 64, pp.\ 3117-3128.

\ref[Ber19b] Bertsekas, D.\ P., 2019.\ Reinforcement Learning and Optimal Control, Athena Scientific, Belmont, MA.

\ref[Ber19c] Bertsekas, D.\ P., 2019.\ ``Robust Shortest Path Planning and Semicontractive Dynamic Programming," Naval Research Logistics, Vol.\ 66, pp.\ 15-37.

\ref[Ber20] Bertsekas, D.\ P., 2020.\
Rollout, Policy Iteration, and Distributed Reinforcement Learning, Athena Scientific, Belmont, MA.

\ref[Ber21] Bertsekas, D.\ P., 2021.\ ``On-Line Policy Iteration for Infinite Horizon Dynamic Programming," arXiv preprint arXiv:2106.00746.

\ref[DeR79] Denardo, E.\ V., and Rothblum, U.\ G., 1979.\ ``Optimal Stopping, Exponential Utility, and Linear Programming," Math.\ Programming, Vol.\ 16, pp.\ 228-244.

\ref [Den67] Denardo, E.\ V., 1967.\  ``Contraction Mappings in the Theory Underlying
Dynamic Programming," SIAM Review, Vol.\ 9, pp.\ 165-177.

\ref[FiT91] Filar, J.\ A., and Tolwinski, B., 1991.\ ``On the Algorithm of Pollatschek and Avi-ltzhak," in Stochastic Games and Related Topics, Theory and Decision Library, Springer, Vol.\ 7,  pp.\ 59-70.

\ref [FiV97] Filar, J., and Vrieze, K., 1997.\ Competitive Markov Decision Processes, Springer,  N.\ Y.

\ref[HoK66] Hoffman, A.\ J., and Karp, R.\ M., 1966.\ ``On Nonterminating Stochastic Games," Management Science, Vol.\ 12, pp.\ 359-370.

\ref[Jac73] Jacobson, D.\ H., 1973.\ ``Optimal Stochastic Linear Systems with Exponential Performance
Criteria and their Relation to Deterministic Differential Games," IEEE Trans.\ Automatic Control, Vol.\ AC-18, pp.\ 124-131.

\ref[Kal20] Kallenberg, L., 2020.\ Markov Decision Processes, Lecture Notes, University of Leiden.

\ref [Kle68] Kleinman, D.\ L., 1968.\  ``On an Iterative Technique for Riccati
Equation Computations," IEEE Trans.\ Automatic  Control, Vol.\ AC-13, pp.\ 114-115.

\ref[Kuh53] Kuhn, H.\ W., 1953.\ ``Extensive Games and the Problem of Information," in Kuhn, H.\ W., and
Tucker, A.\ W.\  (eds.), Contributions to the Theory of Games, Vol. II, Annals of Mathematical
Studies No. 28, Princeton University Press, pp.\ 193-216.

\old{
\ref[Lit96] Littman, M.\ L., 1996.\
``Algorithms for Sequential Decision Making, Ph.D.\ Thesis, Brown University, Providence, RI.
}

\ref[PPG16] Perolat, J., Piot, B., Geist, M., Scherrer, B., and Pietquin, O., 2016.\ ``Softened Approximate Policy Iteration for Markov Games," in Proc.\ International Conference on Machine Learning, pp.\ 1860-1868.

\ref[PSP15] Perolat, J., Scherrer, B., Piot, B., and Pietquin, O., 2015.\ ``Approximate Dynamic Programming for Two-Player Zero-Sum Markov Games," in Proc.\ International Conference on Machine Learning, pp. 1321-1329.

\ref[PaB99] Patek, S.\ D., and Bertsekas, D.\ P., 1999.\ ``Stochastic Shortest Path Games," SIAM J.\ on Control and Optimization, Vol.\ 37, pp.\ 804-824.

\ref[Pat01] Patek, S.\ D., 2001.\ ``On Terminating Markov Decision Processes with a Risk Averse Objective Function," Automatica, Vol.\ 37, pp.\ 1379-1386.

\ref [PoA69] Pollatschek, M., and Avi-Itzhak, B., 1969.\ ``Algorithms for Stochastic
Games with Geometrical Interpretation," Management Science, Vol.\ 15, pp.\ 399-413.

\ref [PuB78] Puterman, M.\ L., and Brumelle, S.\ L., 1978.\  ``The Analytic Theory of
Policy Iteration," in Dynamic Programming and Its Applications, M.\ L.\ Puterman
(ed.), Academic Press, N.\ Y.

\ref [PuB79] Puterman, M.\ L., and Brumelle, S.\ L., 1979.\  ``On the Convergence of Policy Iteration in Stationary Dynamic Programming," Math.\ of Operations Research, Vol.\ 4, pp.\ 60-69.

\ref[Rot84] Rothblum, U.\ G., 1984.\ ``Multiplicative Markov Decision Chains," Math.\ of OR, Vol.\ 9, pp.\ 6-24.

\ref[Sha53] Shapley, L.\ S., 1953.\ ``Stochastic Games," Proc.\ of the National Academy of Sciences, Vol.\ 39, pp.\ 1095-1100.

\ref[Tol89] Tolwinski, B., 1989.\ ``Newton-Type Methods for Stochastic Games," in Basar T.\ S., and Bernhard P.\ (eds),  Differential Games and Applications, Lecture Notes in Control and Information Sciences, vol.\ 119, Springer,  pp.\ 128-144. 

\ref[Van78] van der Wal, J., 1978.\ ``Discounted Markov Games: Generalized Policy Iteration Method," J.\ of Optimization Theory and Applications, Vol.\ 25, pp.\ 125-138.

\ref[Whi81] Whittle, P., 1981.\ ``Risk-Sensitive Linear/Quadratic/Gaussian Control," Advances in Applied Probability, Vol.\ 13, pp.\ 764-777.

 \ref[YuB13] Yu, H., and Bertsekas, D.\ P., 2013.\ ``Q-Learning and Policy Iteration Algorithms for Stochastic Shortest Path Problems," Annals of Operations Research, Vol.\ 208, pp.\ 95-132.

 \ref[YuB15]  Yu, H.,  and Bertsekas, D.\ P., 2015.\ ``A Mixed Value and Policy Iteration Method for Stochastic Control with Universally Measurable Policies," Math.\ of  Operations Research, Vol.\ 40, pp.\ 926-968.

 \ref[Yu14] Yu, H., 2014.\ ``Stochastic Shortest Path Games and Q-Learning," arXiv preprint arXiv:1412.8570.
 
 \ref[ZYB21] Zhang, K., Yang, Z. and Basar, T., 2021.\ ``Multi-Agent Reinforcement Learning: A Selective Overview of Theories and Algorithms.," Handbook of Reinforcement Learning and Control, pp.\ 321-384.

\ref[Zac64] Zachrisson, L.\ E., 1964.\ ``Markov Games," in Advances in Game Theory, M.\ Dresher, L.\ S.\ Shapley, and A.\ W.\ Tucker, Princeton University Press, Princeton, N.\ J., pp.\ 211-253.

\end